\documentclass[final,3p,times]{elsarticle}

\usepackage{graphics,epstopdf,amssymb}
\biboptions{sort&compress}

\newcommand*{\ket}[1]{\ensuremath{\left|{#1}\right\rangle}}
\newcommand*{\etal}[0]{\emph{et~al.}}
\newcommand*{\Rb}[1]{\ensuremath{^{#1}}Rb}

\journal{Physics Reports}

\begin{document}

\begin{frontmatter}

\title{Atom lasers: production, properties and prospects for precision inertial measurement}

\author{N.~P. Robins, P.~A. Altin, J.~E. Debs, and J.~D. Close}

\address{Department of Quantum Science, The Research School of Physics and Engineering, \linebreak The Australian National University, Canberra, Australia, 0200}

\begin{abstract}
We review experimental progress on atom lasers out-coupled from Bose-Einstein condensates, and consider the properties of such beams in the context of precision inertial sensing.
The atom laser is the matter-wave analog of the optical laser. Both devices rely on Bose-enhanced scattering to produce a macroscopically populated trapped mode that is output-coupled to produce an intense beam. In both cases, the beams often display highly desirable properties such as low divergence, high spectral flux and a simple spatial mode that make them useful in practical applications, as well as the potential to perform measurements at or below the quantum projection noise limit. Both devices display similar second-order correlations that differ from thermal sources. Because of these properties, atom lasers are a promising source for application to precision inertial measurements. 
\end{abstract}


\end{frontmatter}

\tableofcontents

\section{Introduction and overview}

The optical laser is now ubiquitous in everyday life, although for more than ten years after Theodore Maiman first demonstrated a laser in 1960 it saw little application. From the outset, the device was clearly intriguing from the point of view of fundamental physics. It was also understood that the laser had potential in the areas of precision measurement and communications, but its failure to find broad application in its first decade prompted the description ``a solution looking for a problem" \cite{C.Townes:2010uq}. In 2001, Professors Eric Cornell, Carl Weiman and Wolfgang Ketterle were awarded the Nobel prize for Bose-Einstein condensation. In his Nobel lecture, Professor Ketterle described a Bose-Einstein condensate as the ``the creation of laser-like atoms". Fourteen years later, the atom laser -- the atomic analogue of an optical laser -- has not yet found a viable application, a situation reminiscent of the optical laser in the early 1970s. That may well be set to change in the near future.

One of the most significant impacts of the optical laser has been in revolutionizing the field of metrology, with lasers now being used to conduct precise measurements in virtually every field, from biology, chemistry and medicine to geology and astronomy, as well as in fundamental physics. Precision metrology has also benefited from advances in techniques for manipulating atoms, which led among other things to the development of atomic clocks based on Ramsey interference \cite{Borde:2002,Wynands:2005}, which now provide a worldwide time and frequency standard and have enabled technologies such as the Global Positioning System (GPS). Compared with photons, atoms have numerous properties which enable them to interact with their environment in ways that light cannot. This makes atoms versatile probes of the physical world, and therefore valuable for metrology. Over the past few decades, precision measurements using atomic sources has come to the fore in many areas of fundamental and applied physics. Atom interferometric measurements of the Newtonian constant of gravitation $G$ \cite{Fixler:2007,Lamporesi:2008} and the fine structure constant $\alpha$ \cite{Gupta:2002,Bouchendira:2011} now reach the precision of the CODATA recommended values, and proposals for tests of general relativity \cite{Dimopoulos:2007,Hohensee:2011} and gravitational wave detection \cite{Dimopoulos:2008a} have gained recent interest. Atom interferometers can also be extremely precise inertial sensors, measuring linear accelerations \cite{Canuel:2006}, rotations \cite{Lenef:1997,Gustavson:1997}, local gravity \cite{Kasevich:1992,Peters:1999} or gravity gradients \cite{Snadden:1998,McGuirk:2002}, with applications in navigation and geophysics. Of course, this versatility is also one of the major drawbacks of atom metrology: the susceptibility of atoms to various physical effects makes it technically challenging to isolate the quantity that is to be measured from other effects (`noise'). However, in many cases the diversity in atomic properties such as mass, magnetic moment, polarisability, interaction strength (both with light and with other atoms) and internal structure allows one to choose the probe best suited to a particular measurement.

In the same way that optical lasers superseded thermal light sources in precision metrology, one might expect that the atom laser -- a source of coherent matter-waves -- may prove to be advantageous in precision measurements with atoms. Interestingly, although the stimulated emission process on which a laser is based is quintessentially quantum-mechanical, it is nearly always the \emph{classical} properties of optical lasers -- their brightness, divergence, spatial mode structure and classical frequency and amplitude stability -- that make them preferred to thermal light sources in many applications. Atom lasers share many of these coveted properties.

\subsection{What is an atom laser?}

To discuss the properties and potential of these devices, we must first define what we mean by an \emph{atom laser}. A detailed definition is given by Wiseman in Ref.\ \cite{Wiseman:1997uq}, taking the view that an essential part of the definition is that a laser is a device with an output that is well approximated by a classical wave of fixed amplitude and phase. As the author notes, his definition precludes devices such as pulsed lasers from being defined as a laser. For our purposes, such a definition is too restrictive. 

In this paper, we take a practical and deliberately broad point of view. For our purposes, a laser is a device that produces a guided or freely propagating beam of bosons that have been out-coupled from one or more macroscopically-occupied trapped modes. In addition to continuous wave (cw) single mode devices, this description encompasses pulsed lasers, mode-locked and Q-switched lasers, multimode cw lasers, super-continuum lasers and most other sources that have become known as optical lasers. Of course it is possible to criticise our rather loose definition. Our statement lacks the tightness of Wiseman's definition, but has the advantage that it captures the essence of most optical devices that we call lasers.

\begin{figure}[t!]
\centerline{\scalebox{.25}{\includegraphics{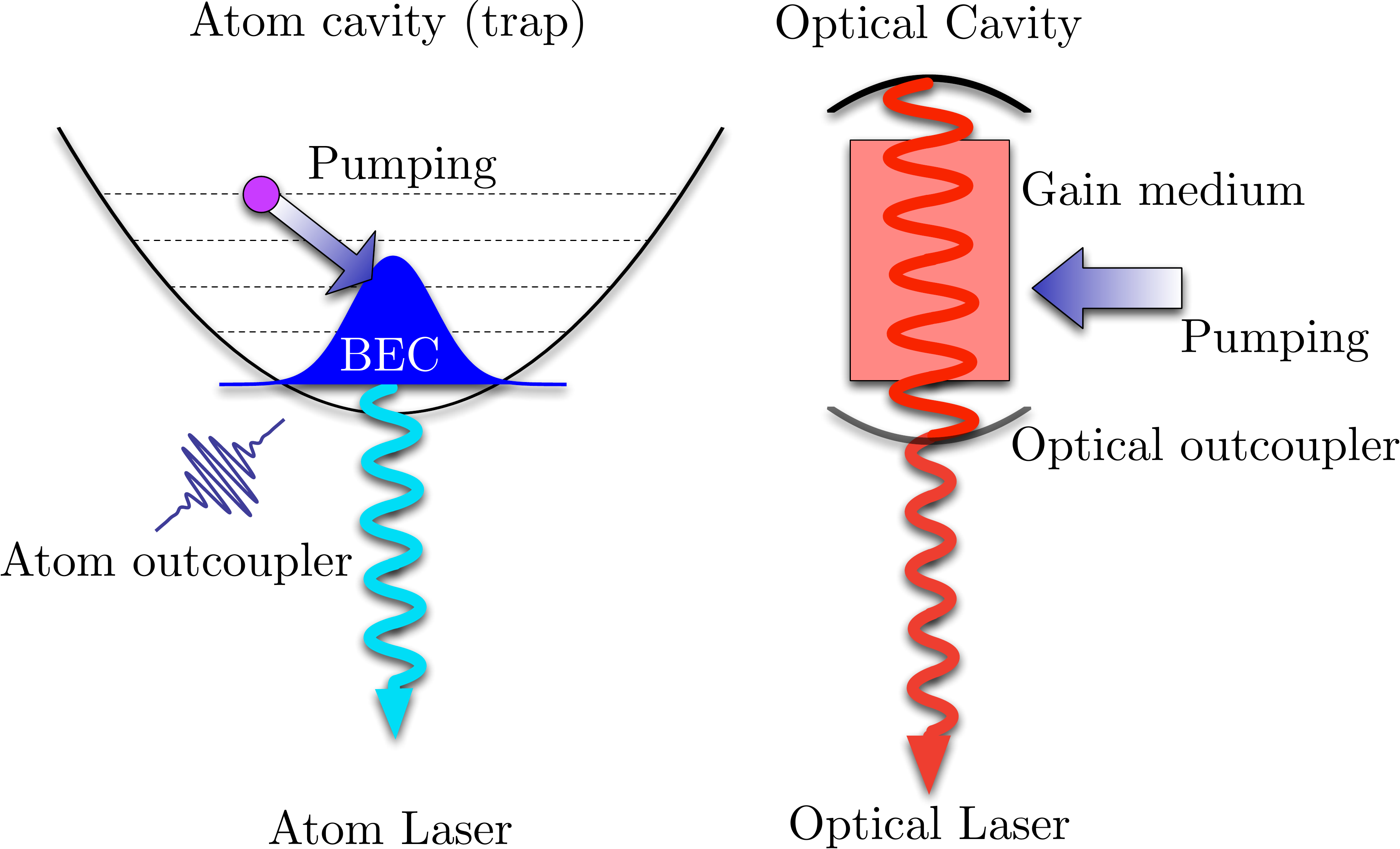}}}
\caption{Comparison of an optical laser and an atom laser. Both devices employ a trap to confine the lasing mode that is macroscopically populated via Bose-enhanced scattering. Because photons do not scatter off other photons in vacuum, an optical laser must employ a nonlinear gain medium to mediate Bose-enhanced scattering. For an optical laser, the trap is formed from material mirrors and/or gratings, and an electromagnetic mode is macroscopically populated. In an atom laser, the roles are reversed: an electromagnetic trap confines a macroscopically trapped matter-wave. Both devices employ an output coupler to couple the lasing mode to the continuum of modes external to the trap.}\label{atlas}
\end{figure}

From a mechanistic point of view, there is great similarity between the operation of optical lasers and atom lasers. A schematic comparison of the two is given in Figure \ref{atlas}. Both devices typically employ a trap to confine the lasing mode. In an optical laser, the trap is the cavity formed by mirrors, gratings or other material elements. In an atom laser, the cavity is an atom trap formed by some combination of electromagnetic fields. Both devices exploit Bose-stimulated or Bose-enhanced scattering to produce a macroscopically-occupied trapped mode and both employ an output-coupler to couple particles from the trapped mode to a propagating beam. In an optical laser, output-coupling is often achieved by making one or both cavity mirrors partially transmitting, or by using a diffraction grating. In an atom laser, output-couplers for magnetically trapped atoms can be realised using radio-frequency (rf) coupling or optical Raman transitions to couple a trapped atomic state to an un-trapped state. Output-coupling from optical traps can be effected by reducing the trapping potential until atoms spill from the trap. In nearly all cases the directionality of the atomic beam is provided by gravity; the exception is Raman out-coupling, in which atoms receive a momentum kick from the Raman beams. These different atom laser output-coupling mechanisms are described and compared in detail in Section \ref{out-couplers}.

%%%%%
%%%%%

\subsection{Atom lasers and precision measurement}

One promising potential application of the atom laser is in the precision measurement of inertial effects such as accelerations and rotations. Over the last decade there has been significant improvement in the stability, size and reliability of cold-atom inertial sensors. Despite this, published sensitivities have not improved substantially over that time. The devices are limited by classical effects such as vibrations \cite{Peters:2001aa}, wavefront aberrations in the optical probe lasers, and the Coriolis effect \cite{Fils:2005aa,Louchet-Chauvet:2011aa,Bouchendira:2011aa}. It is here that the atom laser may well find application. Inertial measurements are highly sensitive to the spatial properties of the cold atom source and the ability to mode match the cold atom source to the optical ruler that forms the beam splitters and mirrors in atom interferometers. The high brightness and low divergence of atom laser sources may well allow us to improve substantially on current cold atom inertial sensing technology. 

In principle, the photon flux determines the quantum noise (or shot noise) limit of precision in any optical measurement. A 1000 Watt light bulb can be purchased for a few hundred dollars. However, the Advanced Laser Interferometer Gravitational Wave Observatory (Advanced LIGO), the most advanced inertial sensor yet created, uses a 200 Watt, highly stabilised Nd:YAG laser as its source \cite{Seifert:2006fk}. This is because, in \emph{practice}, it is often the classical properties of lasers -- brightness, divergence, spatial mode structure and classical frequency and amplitude fluctuations -- that make it possible to reach the quantum noise limit at high flux. Input flux is not everything, and optical lasers with fluxes orders of magnitude below thermal sources are standard in precision optical measurements.

All high-precision cold-atom-based inertial measurements made to date have used laser-cooled thermal atomic sources -- the equivalent of an atomic ``light bulb''. Thermal sources offer a variety of advantages. They typically have higher flux, arguably lower experimental complexity, and lower density at a given flux, alleviating dephasing driven by atomic interactions. They have the disadvantages of high divergence and low brightness in comparison with state-of-the-art atom lasers. \emph{Can atom lasers be developed to the point where their advantages outweigh the disadvantages so they supersede thermal sources in inertial measurements?} This is a question worthy of consideration, and a question that has motivated much of the development of the atom laser.

Precision measurements of accelerations are typically made with respect to an inertial frame provided in a gravitational field by freely falling or approximately freely falling test masses. In the case under discussion here, the test masses are a beam or pulse of falling cold atoms. If the lab frame accelerates while the atoms are in free fall, the location of the atoms will change, as measured by a ruler tied to the accelerating frame. This is the principle of cold atom inertial measurement. The rulers that are used in all state-of-the-art cold-atom inertial sensors are realized with moving-standing waves of light formed from two counter-propagating lasers tuned to drive Raman or Bragg transitions in the atoms. It is critical that the optical lasers that form the ruler are stable in frequency, phase, amplitude, and pointing and that their phase fronts are flat over a spatial region that is on the order of the size of the atomic cloud. Imperfections of the optical ruler over this length scale can limit the sensitivity of the measurement.
 
The quality required of the ruler depends on the spatial properties of the atomic source. A physically small, low divergence beam of atoms will sample a small region of the optical beam profile at the beam splitter and mirror pulses. The optical phase fronts must be flat, ideally to within the atomic shot noise limit of a phase measurement, over this region. Atom lasers have transverse modes with a size and divergence at least two orders of magnitude lower than the thermal atomic sources that are traditionally used in inertial sensors. This is likely to alleviate the impact of imperfections in the optical rulers. The spatial mode of atom lasers and their divergence and brightness properties are discussed in Section \ref{sect:spatialmode} of this review.

A naive comparison of thermal sources and atom lasers would immediately suggest that thermal sources are better suited to precision inertial measurements because their flux is orders of magnitude higher than condensed sources. For example, it is straightforward to produce a cold atom beam with a flux exceeding $10^{11}$ atoms/second using laser cooling alone in a Zeeman slower or two-dimensional magneto-optic trap \cite{Slowe:2005,Klempt:2011aa}. The best BEC sources are capable of just over $10^6$ atoms/second \cite{Stam:2007aa}. However, it is not the total flux entering an interferometer that determines the signal-to-noise ratio, but rather the flux contributing to the signal. Taking into account their reduced mode size and divergence, the useful or \emph{effective} flux of an atom laser can be comparable to that of a thermal beam. The flux limits of atom lasers are discussed and compared to thermal sources in Section \ref{sect:flux}.

\subsection{Coherence and squeezing}

Although the classical properties of lasers are often of great importance in practical applications, it is their quantum coherence properties that strictly distinguish laser sources from thermal sources. Various experiments over the past decade have confirmed that the analogy between optical and atom lasers extends to their higher-order coherence, with atom lasers exhibiting unity-valued correlation functions up to at least third order. The spatial and temporal coherence properties of atom lasers are discussed in Section \ref{sect:coherence}.

The precision of a phase measurement made using a coherent state is limited by quantum uncertainty to the projection noise limit $\Delta\phi = 1/\sqrt{N}$, where $N$ is the mean particle number. It has been known for more than 30 years that this limit may be overcome by squeezing the uncertainty of the quantum state \cite{Caves:1981}. Squeezing the quantum noise on optical lasers has been an active field of research for more than 20 years, and the field has produced many significant results \cite{Kolobov:1999aa,Reid:2009aa}. In 2011 the gravitational wave detector GEO600 demonstrate improved interferometric sensitivity through the injection of squeezed vacuum \cite{Collaboration:2011aa}.

Nevertheless, there is an important difference between photons and atoms which makes it likely that squeezing will prove to be even more important in the development of atom lasers for applications in metrology. Coherent photons are cheap and easy to produce, and it is nearly always simpler to increase the photon flux rather than to employ squeezing to improve the precision of a measurement. Cold atoms are presently much more difficult amd expensive to produce. A laser pointer that can be bought for a few dollars produces $10^{16}$ photons per second and fits in the palm of your hand. A cold atom source appropriate for precision measurement currently costs perhaps five orders of magnitude more, has a flux $10^8$ times lower and is 1000 times larger. Further increases to the flux of atomic sources will come only at considerable expense and difficulty. With this restriction, squeezing becomes a viable candidate for improving the precision of atom-laser-based measurements. High quantum efficiency atom detection, necessary to exploit the promise of squeezing, has already been demonstrated by several groups \cite{Ottl:2006aa,Teper:2006aa,Heine:2010aa,Bakr:2010aa,Sherson:2010aa,KohnenM.:2011aa,poldy:013640,Takamizawa:2006aa}.

There are other aspects that make the prospects for exploiting squeezing in practical measurements using atoms look rather more promising. Squeezing proceeds via interactions or nonlinear terms in the Hamiltonian. Photons do not interact with photons in vacuum and require a substantial engineering effort to produce a non-linear medium capable of substantial squeezing. Atoms, on the other hand, interact with each other and squeeze ``for free''. In addition, optical squeezing is very susceptible to loss due to scattering and absorption in a typical lab environment. Cold atom experiments by necessity are already performed in an extremely low-loss, clean vacuum environment. Squeezing of atom laser beams is discussed further in Section \ref{sect:squeezing}.

\subsection{Pumping an atom laser}

The operation of atom lasers has thus far been restricted to either the pulsed or the quasi-continuous regime by depletion of the source Bose-Einstein condensate. A truly continuous, phase-coherent atom laser has yet to be achieved, although a number of major steps have been made towards such a system. In the context of metrological applications, pulsed measurements limit the bandwidth of signals that can be detected and additionally alias high frequency noise into the signal band (the Dick effect \cite{Dick:1987}). A continuous atom laser would extend the utility of atom-based inertial sensors to much higher bandwidth measurements.

To realize a continuous atom laser, the source Bose-Einstein condensate must be replenished or `pumped' in a sustained and phase-coherent fashion. The pumping mechanism must be irreversible in order to achieve a net transfer of atoms into the lasing mode. Experimental and theoretical work on possible pumping mechanisms is reviewed in Section \ref{pumping}.

\subsection{Experimental demonstrations of inertial measurements with atom lasers}

In the final sections, we review experimental progress on inertial measurements already made using atom lasers, and compare the performance of atom laser and thermal sources in inertial sensors. The discussion focuses on the issues of interaction-induced dephasing and the role of momentum width. We conclude with a discussion of the future prospects for atom lasers in precision measurement.

\section{Atom laser output-coupling mechanisms} \label{out-couplers}

The collection of atoms into a macroscopically-occupied trapped mode (Bose-Einstein condensation) has been discussed in great detail in many reviews and textbooks (see, for example, Refs.\ \cite{Dalfovo:1999,Pethick:2002}), and will not be covered here. Once the source is prepared, all that remains is to extract atoms from the condensate into a guided or freely propagating beam. Exactly how this is done is crucial in determining the properties of the atom laser, in particular its spatial mode, divergence and brightness.

Five methods have been explored to couple atoms from a trapped Bose-Einstein condensate into an atom laser beam: rf and Raman output-coupling, Majorana spin flipping in a magnetic trap, Landau-Zener sweeps and spilling/tunnelling from an optical trap. rf coupling, Raman coupling and Majorana coupling all operate in a similar way, by coupling atoms from a magnetically trapped state to an untrapped state. The untrapped state falls away from the trapped condensate under the influence of gravity and any mean-field repulsion arising from atom-atom interactions. Raman coupling can provide an additional momentum kick to remove atoms from the trapping region more quickly. It is also possible to lower the edge of the trapping potential until it becomes too weak to support the condensate. In a far-detuned optical trap, this is easily achieved by lowering the trap power; atoms then tunnel or spill out of the potential and form an atom laser beam.

\subsection{rf output-couplers}

Most of the work done on atom lasers has used rf out-coupling of atoms from a magnetic trap. In this technique, rf radiation is used to drive magnetic dipole transitions between neighboring Zeeman sublevels, coupling a trapped  state to an $m_F = 0$ untrapped state. The rf radiation is applied by driving an oscillating current in a coil located near the condensate with the axis of the coil perpendicular to the trap bias field. The $m_F = 0$ state populated by the rf coupling pulse forms the atom laser beam. This state exhibits no first order Zeeman shift, and the atoms fall away from the trapped condensate due to gravity.

\begin{figure}[t!]
\centerline{\scalebox{.6}{\includegraphics{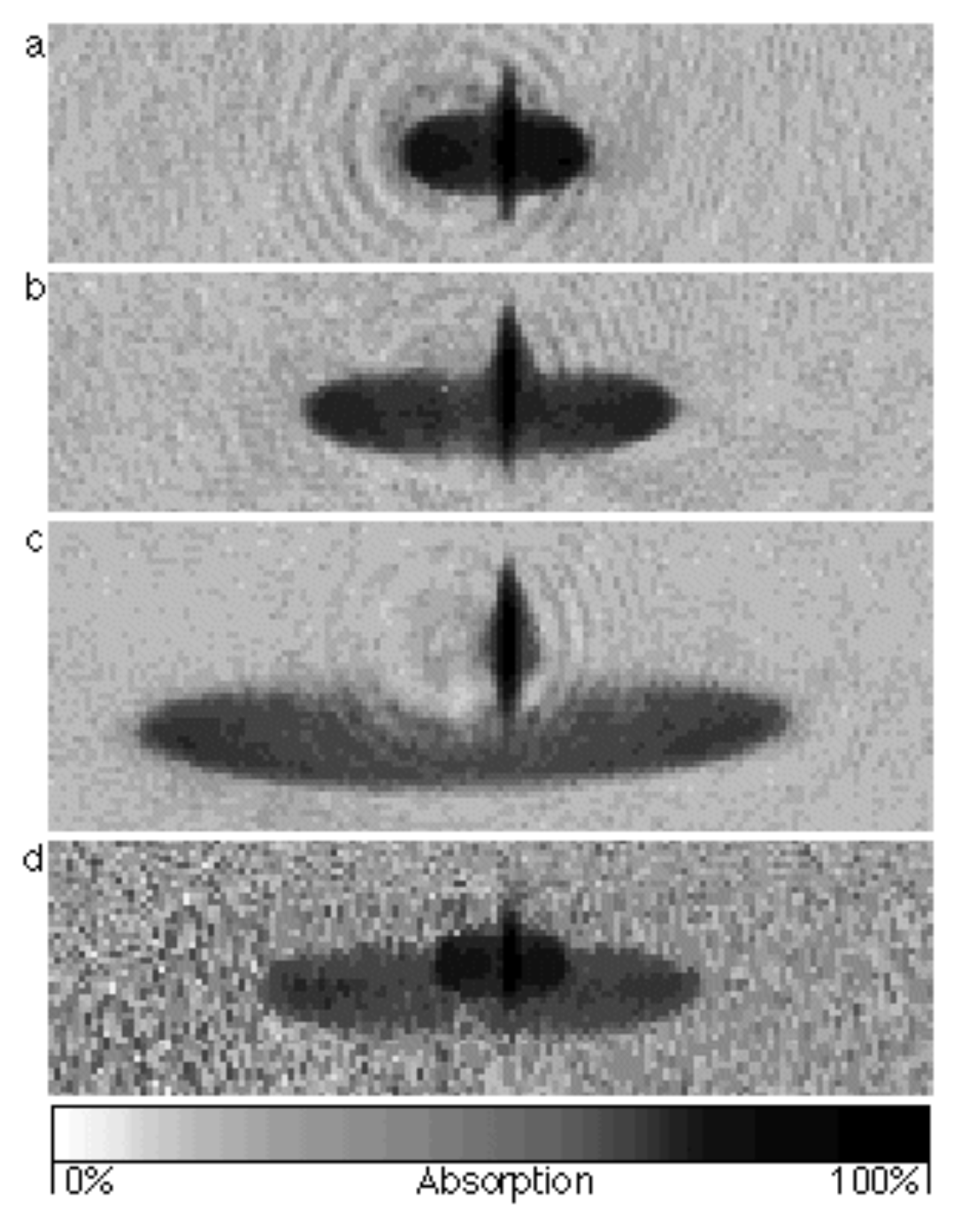}}}
\caption{The first demonstration of out-coupling from a Bose-Einstein condensate using radiofrequency. Absorption images of atoms coupled to the untrapped \ket{m_F= 0} state recorded (a) 14\,ms, (b) 20\,ms, and (c) 25\,ms after a short rf pulse. The trapped condensate appears as a thin line in the centre of each image. (d) Two pulses of atoms coupled out of the same condensate by consecutive rf pulses spaced 10\,ms apart. The image was taken 10\,ms after the second pulse, and has a noisier background due to lower probe laser power. The images are 3.1\,mm across, and were taken along the direction of gravity. Reproduced from \cite{Mewes:1997aa} with permission of the authors.}\label{ketterleAL}
\end{figure}

The first demonstration of an atom laser used rf out-coupling and was performed by the group at MIT \cite{Mewes:1997aa}. In this pioneering work, $5\times10^6$ Bose-condensed sodium atoms in the $\ket{F=1,\,m_F =- 1}$ hyperfine state were confined in a magnetic trap and comprised the lasing atomic mode. A short pulse of rf resonant with the Zeeman splitting at the bottom of the trap was used to couple atoms to the $m_F= 0$ untrapped state. Absorption images of the atoms falling out of the condensate are shown in Figure \ref{ketterleAL}.

%%%%%
%%%%%

Because of its spatial extent in the magnetic trap, a magnetically-trapped condensate extends over a range of field strength, and therefore Zeeman splittings. The MIT work was performed using the short pulse limit where the Fourier width of the coupling pulse was sufficient to address the entire condensate, providing spatially independent coupling. output-coupling in this regime produces copies of the condensate where the number of atoms in the out-coupled, untrapped-state is determined by the duration and amplitude of the rf pulse. The MIT group investigated the number of output-coupled atoms and demonstrated Rabi oscillations as a function of the amplitude of the rf pulse. They also investigated Landau-Zener sweeps of the rf frequency and Majorana coupling. However, although these methods are interesting in their own right, it was unswept rf coupling that became the out-coupler of choice in many subsequent experiments \cite{Bloch:1999aa,Martin:1999ys,Le-Coq:2001aa,Robins:2004fk,Dall:2007aa}.

The advantage of short out-coupling pulses is that the stability requirements on the magnetic bias field and on the frequency of the rf coupling field are relaxed, due the the broad Fourier width of the pulse. Any change of bias field or drift in the resonance frequency has a rather minimal effect unless it is on the order of the Fourier width of the pulse. Short rf pulses produce a lower density copy of the lasing condensate with (at least initially), the same spatial distribution and momentum distribution. As the untrapped atom pulse falls away from the condensate, its spatial extent increases due to the momentum distribution of the atoms and due to mean field repulsion. The momentum distribution of the falling atom pulse is also influenced by mean field effects. This is particularly pronounced in the initial stages as the out-coupled pulse falls away from the condensate. The condensate is dense and the slowly falling atom pulse rolls off the mean field potential hill created by the condensate. We will return to this effect in some detail when we discuss the spatial mode of atom laser beams. 

\begin{figure}[t!]
\centerline{\scalebox{.4}{\includegraphics{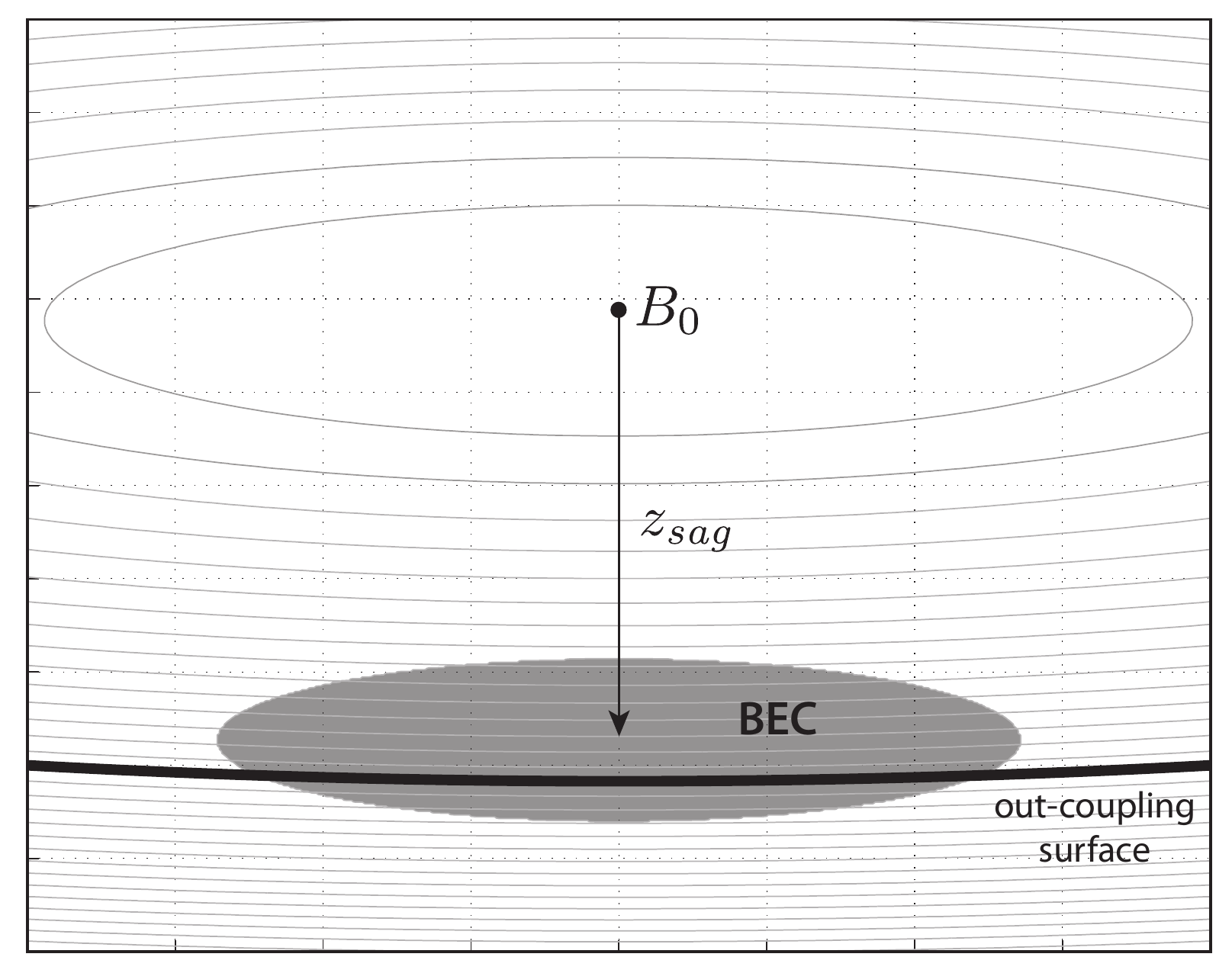}}}
\caption{Gravity plays an important role in determining the out-coupling dynamics of rf and Raman atom lasers. This schematic diagram shows magnetic equipotential surfaces and the equilibrium position of a BEC in a magnetic trap. The condensate is displaced by $z_{sag}$ from the field minimum $B_0$ under the influence of gravity. When the sag is large compared with the size of the condensate, the resonant out-coupling surface is approximately planar.  Figure after Bloch \etal\  \cite{Bloch:1999aa}.}\label{fig:gravsag}
\end{figure}

As the duration of the rf out-coupling pulse increases, its Fourier width decreases and the out-coupling resonance region can become substantially smaller than the condensate itself. In the absence of gravity, the coupling surface is an ellipsoid centred around the condensate. Gravity causes the condensate to sag in the trap, displacing its centre by a distance $z_{sag} = g/\omega_z^2$ below the minimum of the magnetic field, where $g$ is the acceleration due to gravity and $\omega_z$ is the vertical trapping frequency. Under the influence of gravity, the out-coupling surface cuts through the condensate and is no longer a closed surface (Figure \ref{fig:gravsag}). For very weak traps, the surface approaches a plane whose location is determined by the rf output-coupling frequency \cite{Bloch:1999aa,RobinsPhD}. 

%%%%%
%%%%%

The first long pulse (quasi-continuous) rf out-coupled atom laser was realised by Bloch \etal\ in 1999 \cite{Bloch:1999aa}. In this work, a 100\,ms burst of rf radiation was applied to a condensate in the $\ket{F=2,\,m_F = 2}$ state of \Rb{87}, creating a beam of atoms that propagated for several millimetres with divergence below the experimental resolution limit of 3.5\,mrad. In contrast to earlier short pulse experiments, the Fourier width of this output-coupling pulse was an order of magnitude narrower than the resonant width of the condensate. The stability requirements were therefore significantly more stringent than those for short pulse results. This long pulse output-coupler was an important step towards the as yet unrealised goal of a truly continuous atom laser. Bloch \etal\ reported a neutral atom beam with a brightness exceeding $2\times10^{24}$ atoms s$^2$\,m$^{-5}$, a value unprecedented in atom optics. The Fourier limited linewidth and high brightness of quasi-continuous atom lasers makes them promising candidates for precision interferometry, particularly when combined with large momentum transfer (LMT) beam splitters as discussed in Section \ref{sect:precmeas}.

\subsection{Raman output-couplers}

\begin{figure}[t!]
\centerline{\scalebox{.25}{\includegraphics{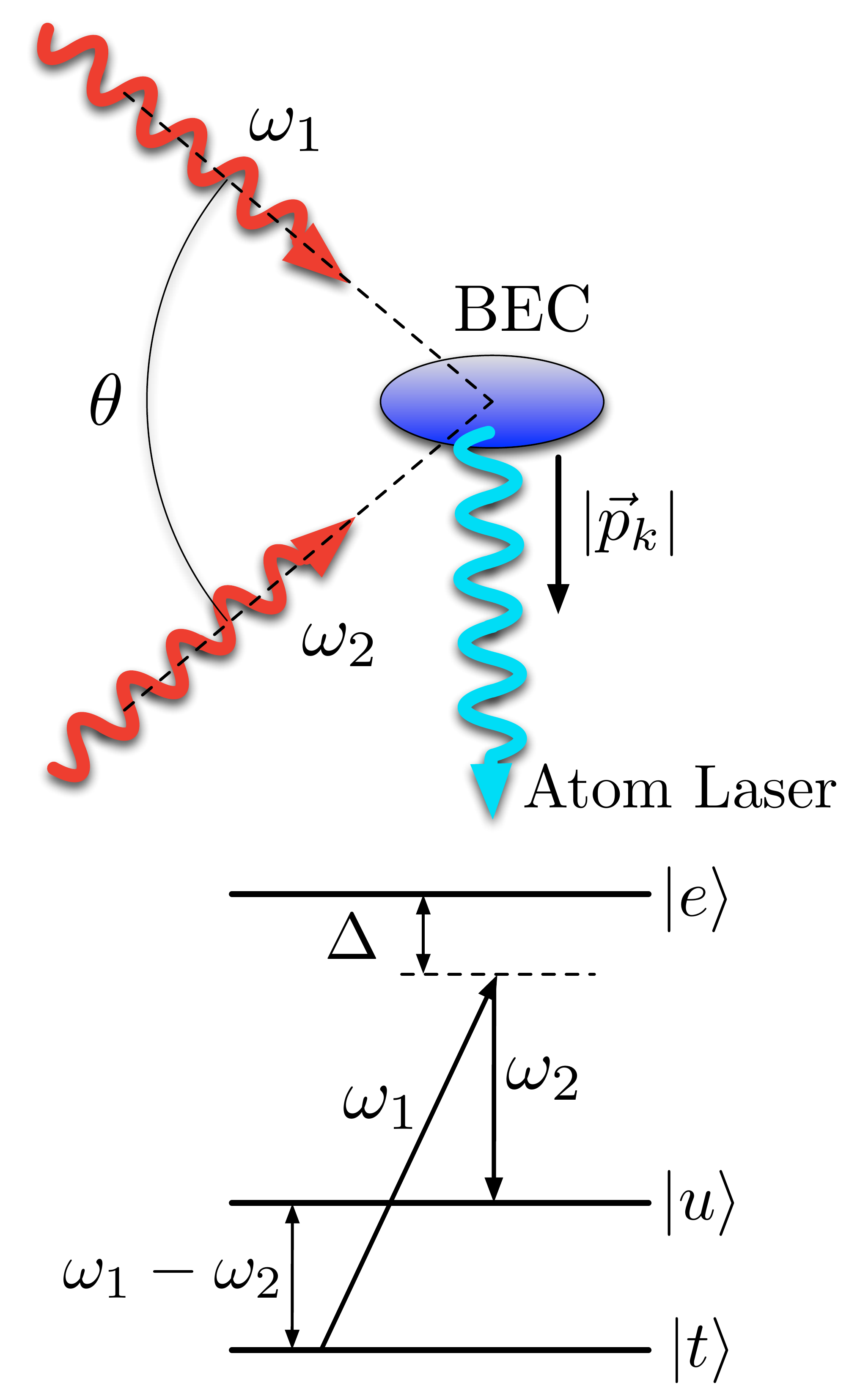}}}
\caption{Two optical beams incident on a condensate drive a two-photon Raman transition from a trapped state to an untrapped state to form an atom laser beam. The frequency difference of the two beams is tuned to match the energy difference of the trapped and untrapped states including the recoil energy of the final state. The untrapped atoms recoil with momentum $p_{rec} = 2\hbar k \sin(\theta/2)$, where $k$ is the optical wavenumber of the Raman beams and $\theta$ the angle between them.}\label{raman}
\end{figure}

The coupling from a trapped to an untrapped state can also be realized via a two-photon Raman transition. This technique also has the potential to provide a significant momentum kick to the atoms as they leave the trap. The process is illustrated in Figure \ref{raman} and involves three atomic levels, the trapped ground state that is the lasing mode, the recoiling untrapped state and an excited state of the atom. Two phase-locked beams with a frequency difference $\hbar\delta = \hbar (\omega_1 - \omega_2)$ are directed onto the condensate. A photon is absorbed from $\omega_1$ and stimulated to emit into $\omega_2$, transferring momentum equal to $\vec k_1 - \vec k_2$, where $\vec k_i$ is the wave vector of the $i$th beam. To satisfy conservation of energy and momentum between the trapped and untrapped (recoiling) state, $\delta = \delta_Z + \omega_{rec}$, where $\hbar\delta_Z$ is the internal state energy resonance, and $\hbar \omega_{rec}$ the kinetic energy associated with the recoil momentum. Although the two frequencies are chosen to satisfy two-photon resonance, individually they are far detuned from the excited state to suppress spontaneous emission. Although Raman out-coupling is more technically demanding than rf out-coupling, imparting a momentum kick can significantly improve the spatial mode of the atom laser beam as well as increasing its flux. These effects will be discussed further in Sections \ref{sect:spatialmode} and \ref{sect:flux}.

%%%%%
%%%%%

\begin{figure}[t!]
\centerline{\scalebox{1}{\includegraphics{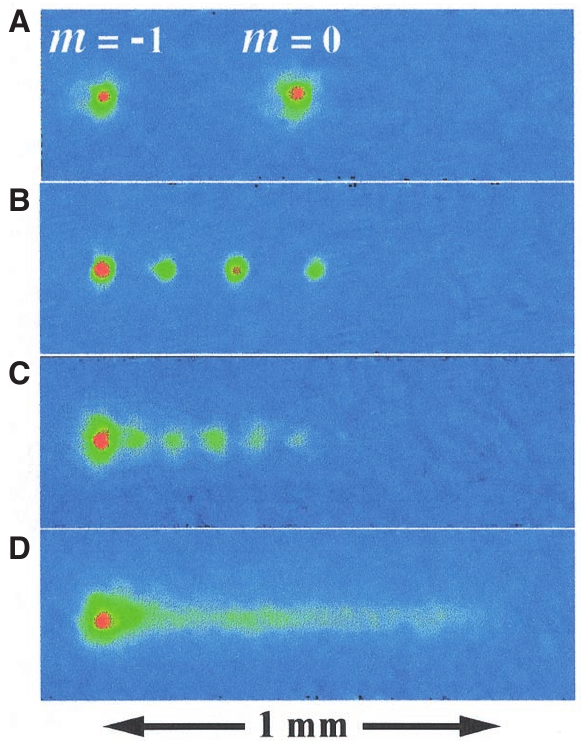}}}
\caption{Raman atom laser operated in pulsed mode. (A to C) One, three, and six 6\,$\mu$s Raman pulses were applied, each out-coupling a copy of the trapped condensate. (D) Firing Raman out-coupling pulses at the full repetition rate of about 20\,kHz (140 pulses in 7 ms) imposed by the frequency of the TOP trap's rotating bias field produces a quasi-continuous atomic beam. Reproduced from \cite{Hagley:1999aa} with permission of the authors.}
\label{fig:hagley}
\end{figure}

The first Raman out-coupled atom laser was demonstrated by Hagley \etal\ in 1999 \cite{Hagley:1999aa}. A magnetically trapped sodium condensate was subjected to short Raman pulses resonant with the $^3S_{1/2},\,F=1 \rightarrow ^3P_{3/2},\,F=2$ transition. The Fourier width of the 6\, $\mu$s coupling pulses vastly exceeded the resonant width of the condensate, and thus the atom laser operated in a similar regime to the short pulse rf out-coupling described earlier. The pulses were applied at a variable repetition rate up to 20\,kHz, creating a quasi-continuous atomic beam (Figure \ref{fig:hagley}). Although the longitudinal momentum width of the atom pulses was not measured in the experiment, it would presumably have been limited by the momentum width of the lasing condensate. The first long pulse Raman atom laser producing a narrow momentum spread beam was performed by Robins and coworkers in 2006 \cite{Robins:2006aa}, and extended to hyperfine state coupling by Debs \etal\ \cite{Debs:2009aa}. Hyperfine state coupling allows the realization of a pure two-level system, which simplifies the atom laser dynamics. Two-state coupling can also be achieved via a single-photon microwave transition between the hyperfine levels \cite{Ottl:2006}, but without the benefit of a large momentum kick.

%%%%%
%%%%%

\subsection{Output-coupling from an optical trap}

Optical dipole traps exploiting the dipole force exerted by a laser on a neutral atom when it is far detuned from the atomic resonance. In an optical trap, all Zeeman states are trapped and the only way to out-couple atoms is by spilling, tunneling or kicking them from the trap. There have been relatively few experiments studying atom lasers derived from optical traps. Anderson and Kasevich produced a pulsed atom laser by loading an \Rb{87} condensate into a vertical, red-detuned standing wave trap \cite{Anderson:1998ly}. The strength of the trap was reduced and atoms tunnelled from a trapped state to a freely propagating state. The spatial phase shift induced by gravity on the vertically aligned array of emitters led to a pulsed output that was amplitude modulated at a frequency $\omega = mg\lambda/2\hbar$, with $\lambda$ the optical wavelength. This process is analogous to Josephson tunnelling.

Cennini \etal\ reported the first experiment where atoms were output-coupled from an optically trapped BEC by reducing the optical potential until the trap could no longer support the atoms against gravity \cite{Cennini:2003aa}. This scheme has also been investigated by Altin \etal\ \cite{Altin:2012aa,Debs:2012aa}. In a conceptually similar experiment, Couvert \etal\ output-coupled atoms from an optical trap into a waveguide by applying a magnetic field gradient. In this experiment, the magnetic field gradient played a role similar to that of gravity in the experiment of Cennini \etal\ \cite{Couvert:2008aa}. We will return to this technique in the next section where we discuss the spatial mode of atom laser beams and compare the mode produced by different output-coupling techniques.

\section{Spatial mode} \label{sect:spatialmode}

A clean output mode is a highly desirable feature in a laser beam, facilitating beam shaping and mode-matching. From a practical perspective, the divergence angle of the beam can be a simple but very useful measure of the quality of the spatial mode. A more rigorous characterisation is provided by the beam quality factor $M^2$, introduced for atom lasers by Riou \etal\ \cite{Riou:2006aa}. As in optics, the $M^2$ value is a measure of how far the beam deviates from the Heisenberg limit \cite{Siegman}. It is defined by
\begin{equation}
M^2 = \frac{2}{\hbar} \Delta x \Delta p_x \,,
\label{M2-def}
\end{equation}
where $\Delta x$ is the beam width measured at the waist, and $\Delta p_x$ is the transverse momentum spread. An ideal (Gaussian) beam would therefore have $M^2 =1$ along both its principal transverse axes.

\begin{figure}[t!]
\centerline{\scalebox{.6}{\includegraphics{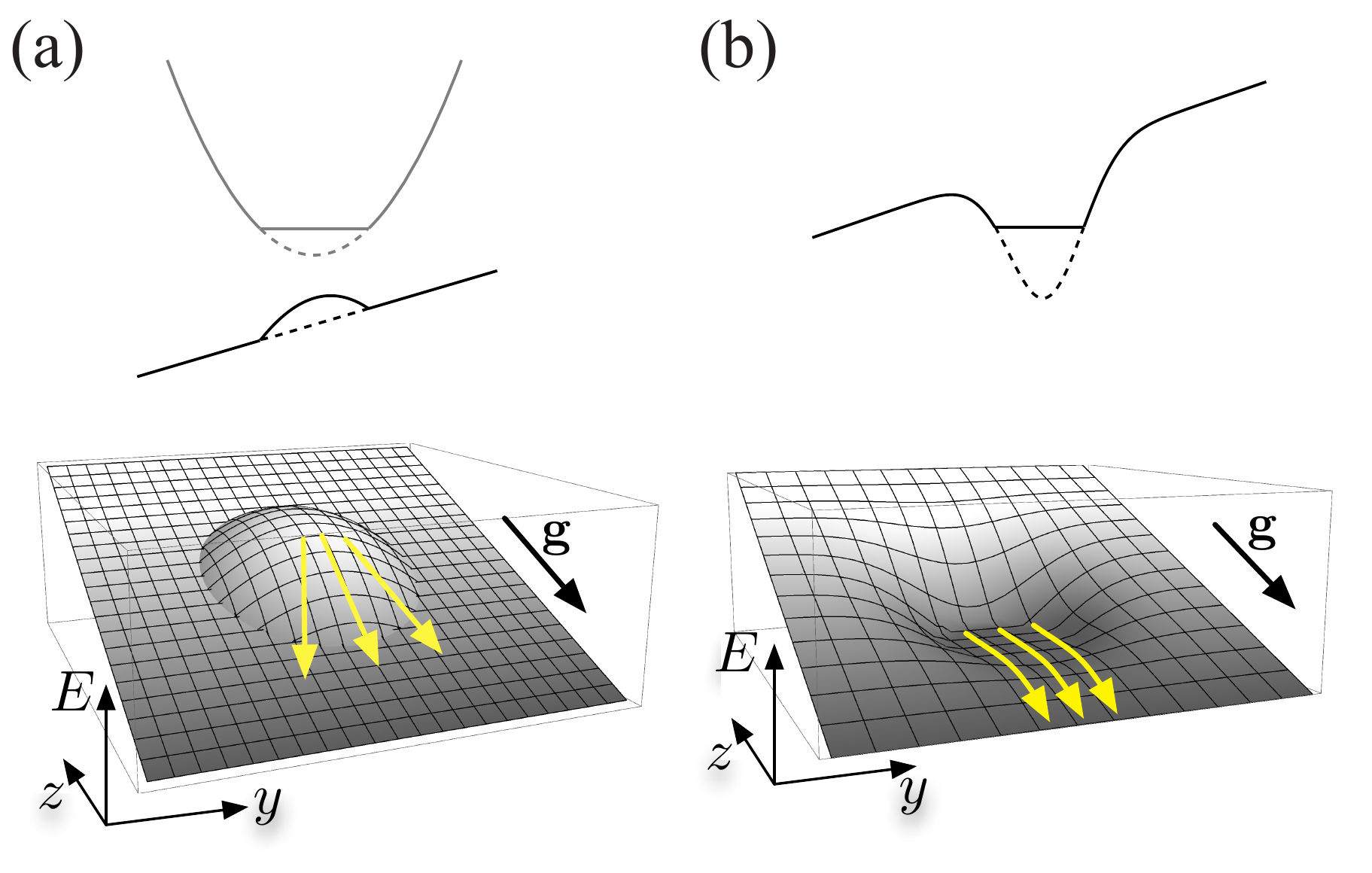}}}
\caption{The impact of interactions on atom laser divergence. (a) When out-coupling from a magnetic trap, atoms in the m$_F$ = 0 state gain transverse momentum as they roll off the potential hill created by the mean-field due to the trapped condensate. (b) In the case of an optical trap, the out-coupled atoms see the same external potential as those in the BEC, and the trapping potential cancels the mean field inside the condensate. Reproduced from \cite{Altin:2012aa} and \cite{Debs:2012aa} with permission of the authors.}
\label{altin-explain}
\end{figure}

It is useful to make some general statements regarding state-changing output-coupling from magnetic potentials before we examine different out-coupling mechanisms individually. The quality of an atom laser beam produced via (rf or Raman) state-changing output-coupling is limited by interactions between the out-coupled atoms and the parent condensate. The reason for this is illustrated in Figure \ref{altin-explain}. For an $m_F = -1$ magnetically trapped condensate in the Thomas-Fermi limit, the mean field interaction between the atoms cancels the trapping potential inside the condensate, shown by the grey curves in the upper panel. Out-coupled atoms in the $m_F=0$ state do not see the trapping potential, but still experience the repulsion from atoms in the condensate. These atoms therefore experience a potential hill, which accelerates them in the direction of gravity and also imparts a transverse velocity dependent on their lateral displacement, causing the beam to diverge. In addition, due to the finite extent of the repulsive potential, atoms emitted from different parts of the condensate may cross in the far field, causing quantum interference that corrupts the spatial mode. The situation is different for atoms which spill or tunnel from an optical potential, as shown in Figure \ref{altin-explain}b.

%%%%%
%%%%%

\subsection{rf output-couplers}

The first study of the spatial mode of an atom laser was made by Le Coq \etal\ \cite{Le-Coq:2001aa}. These authors used long pulse (10ms) rf output-coupling to transfer atoms from the trapped $\ket{F=1,\,m_F=-1}$ state of a \Rb{87} condensate to the untrapped $\ket{F=1,\,m_F=0}$ state. The trap was sufficiently weak that gravitational sag provided a planar output-coupling surface whose location could be adjusted by varying the frequency of the rf coupling field. These authors measured both flux and divergence as function of the position of the output-coupling plane, and analysed the divergence of the beam using an \textit{ABCD} matrix treatment analogous to that applied to the propagation of optical lasers in the paraxial regime. It was found that lensing due to the interaction of the atom laser beam with the condensate was the dominant contribution to the divergence of the beam in that experiment, and that the minimum divergence was observed for output-coupling near the bottom of the condensate. The minimum observed divergence angle was 6\,mrad. 

\begin{figure}[t!]
\centerline{\scalebox{.7}{\includegraphics{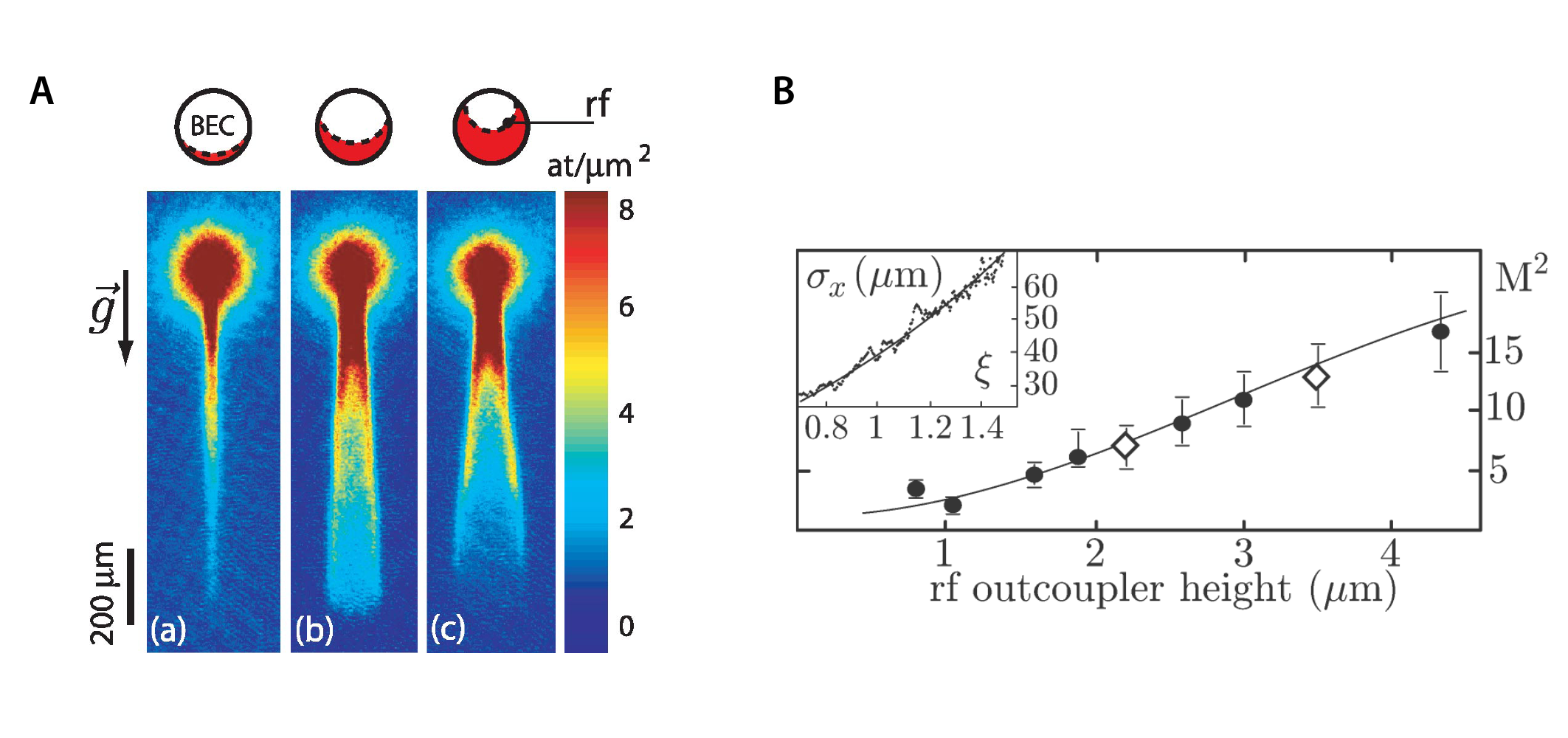}}}
\caption{The effect on the spatial mode of an atom laser of out-coupling from different locations within the source condensate. (a) Absorption images of a non-ideal atom laser, integrated along the elongated axis of the BEC. The figures correspond to different heights of the rf out-coupling surface with respect to the bottom of the BEC: (a) 0.37\,$\mu$m, (b) 2.22\,$\mu$m, (c) 3.55\,$\mu$m. The diagrams above each image show the out-coupling surface (dashed line) and the region of the trapped BEC which is crossed by the atom laser (red). This results in the observation of caustics. The field of view is 350\,$\mu$m $\times$ 1200\,$\mu$m for each image. (b) $M^2$ quality factor vs rf out-coupler distance from the bottom of the BEC: theory (solid line), experimental points (circles). The two diamonds represent the $M^2$ for the two non-ideal atom lasers shown in Figs.\ 1(b) and 1(c). The rf out-coupler position is calibrated by the number of out-coupled atoms. Inset: typical fit of the laser rms size with the generalized Rayleigh formula (Eq.\ 5 in Ref.\ \cite{Riou:2006aa}) for rf out-coupler position 3.55\,$\mu$m. Reproduced from \cite{Riou:2006aa} with permission of the authors.}
\label{figlecoq}
\end{figure}

In 2006, the same group made a more detailed investigation of the spatial mode of an rf output-coupled atom laser and introduced more sophisticated theoretical analysis of the results \cite{Riou:2006aa}. The effect of location of the output-coupling surface is shown clearly in Figure \ref{figlecoq}a. The $M^2$ quality factor for their atom lasers was found to increase as the output-coupling surface approached the centre of the condensate (Figure \ref{figlecoq}b). In the same year, K\"ohl \etal\ performed a closely related study of the divergence of an rf output-coupled atom laser \cite{Kohl:2005aa}. These authors employed a novel curved magnetic mirror to improve the spatial resolution of the transverse structure allowing transverse momentum resolution of 1/60 of a photon recoil, also concluding that the transverse momentum distribution was determined by the mean-field potential of the condensate.

%%%%%
%%%%%

\subsection{Raman output-couplers}

Hagley \etal\ made the first study of a Raman out-coupled atom laser \cite{Hagley:1999aa}. Raman output-coupling removes atoms from the condensate region faster than rf output-coupling due to the two photon momentum kick provided in the output-coupling process, and the mean field potential energy arising from the interaction of the atom laser beam with the lasing condensate is channeled primarily into the forward direction. Although Hagley \etal\ did not make a detailed study of the spatial mode in their work, they discussed the improvement in directionality and the mitigation of mean field effects on the transverse mode structure.

The first detailed comparison of the spatial mode structure of rf and Raman out-coupled atom lasers was performed by Jeppesen \etal\ \cite{Jeppesen:2008aa}. Their experimental setup enabled the production of a high quality atom laser while out-coupling from the center of the condensate rather than at the bottom of the condensate as had been necessary in the rf output-coupled beams discussed earlier. In this work, BECs of $5 \times 10^{5}$ \Rb{87} atoms in the \ket{F=1,\,m_F=-1} state were created via standard runaway evaporation of laser cooled atoms. The apparatus provided the option of transferring atoms to the untrapped \ket{F=1,\,m_F=0} state via either rf or Raman transitions, allowing a comparison of the spatial mode produced using the two techniques in the same apparatus. The system was observed using standard absorption imaging along the weak trapping direction. The rms width of the atom laser as a function of fall distance was extracted from the images and used to calculate the divergence and $M^2$.

\begin{figure}[t!]
\centerline{\scalebox{.6}{\includegraphics{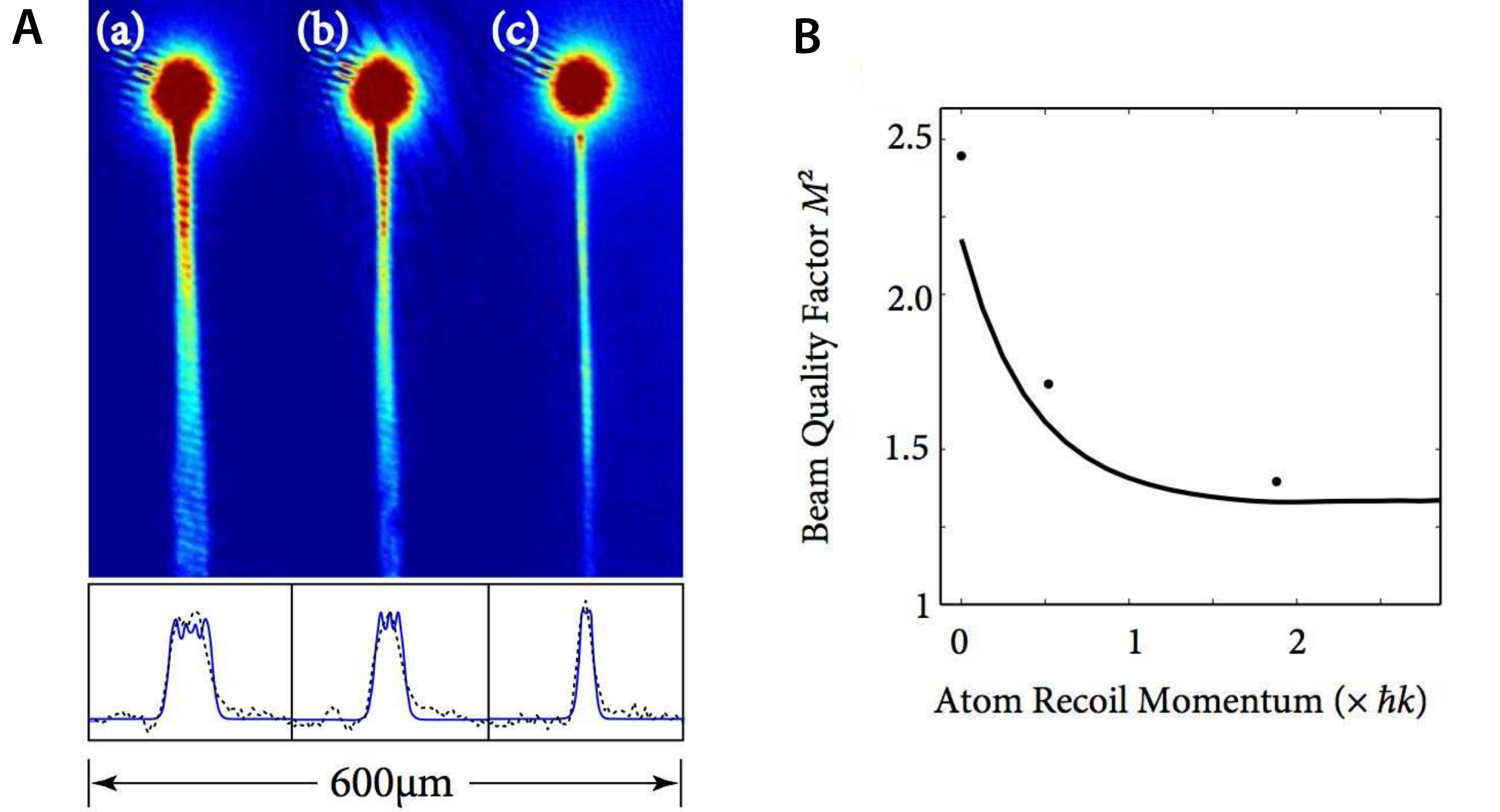}}}
\caption{Improving the spatial mode of an atom laser using Raman output-coupling. (A) Atom laser beams produced using (b) rf and (b, c) Raman transitions. The angle between the Raman beams was $\theta=30^{\circ}$ in (b) and $\theta=140^{\circ}$ in (c), corresponding to a kick of $0.5 \hbar k$ (0.3\,cm/s) and $1.9 \hbar k$ (1.1\,cm/s) respectively. With a larger momentum kick imparted by the Raman beams, the out-coupled atoms obtain less transverse momentum from the mean-field of the trapped condensate. (B) Calculated quality factor $M^2$ of a \Rb{87} atom laser as a function of the initial momentum imparted by the output-coupling process. The dots are experimental values, and the solid lines represent theoretical predictions. The condensate number was $5\times10^5$ atoms, and the ratio of the axial and radial trapping frequencies was 10. Reproduced from \cite{Jeppesen:2008aa} with permission of the authors.}
\label{jeppesen}
\end{figure}

The divergence and $M^2$ value for the Raman atom laser decreased as the momentum imparted by the Raman process was increased (see Figure \ref{jeppesen}). In addition to the experimental results, Jeppesen \etal\ performed a detailed theoretical analysis of the data following the methods introduced by Riou \etal\ \cite{Riou:2006aa}. Inside the condensate, the model was based on the WKB approximation, by integrating the phase along the classical trajectories of atoms moving in the Thomas-Fermi potential of the condensate (an inverted paraboloid). The atom laser wavefunction was propagated outside the condensate using a Kirchoff-Fresnel diffraction integral over the surface of the condensate. Similar to the analysis previously performed by Riou \etal\, the model included only interactions between condensate atoms and beam atoms; interactions between atoms within the beam were ignored. In contrast to the analysis of Riou \etal\, the second order Zeeman shift on the output-coupled beam was not included in the analysis. In both studies, this effect amounted to a very small contribution to the observed divergence.

%%%%%
%%%%%

Experimental and theoretical results from this study are given in Figure \ref{jeppesen}. Figure \ref{jeppesen}a gives example absorption images as the kick is progressively increased, showing a decrease in the divergence of the atom laser. Figure \ref{jeppesen}b gives a comparison between theory and experiment. As the kick was increased, $M^2$ continued to improve, and approached but did not reach the Heisenberg limit of 1, asymptoting to 1.3. In the large kick regime the interaction of the out-coupled atoms with the condensate became negligible and was limited by the non-ideal (non-Gaussian) condensate wavefunction. In addition, the simulations indicated that (using the same maximum two photon kick) it is possible to reach the condensate limit even for much tighter trapping potentials.

Although Raman out-coupling has advantages over rf, it comes with substantially increased technical overhead that may be undesirable in future precision measurement applications. A comparison of the two methods in a state of the art precision inertial measurement is yet to be performed.

\subsection{Output-coupling from an optical trap}

The situation is quite different when atoms are output-coupled from an optical trap. The optical potential appears the same to atoms in any $m_F$ state, and its finite range permits output-coupling without changing the internal state. Atoms in the beam therefore do not see a mean-field potential hill. In the Thomas-Fermi limit, it is cancelled by the trapping potential as illustrated in Figure \ref{altin-explain}b. As a consequence, atoms out-coupled from optical traps can be expected to have much lower divergence and better spatial profile than beams derived from magnetically trapped condensates with the same interaction strength.

%%%%%
\begin{figure}[t!]
\centerline{\scalebox{.5}{\includegraphics{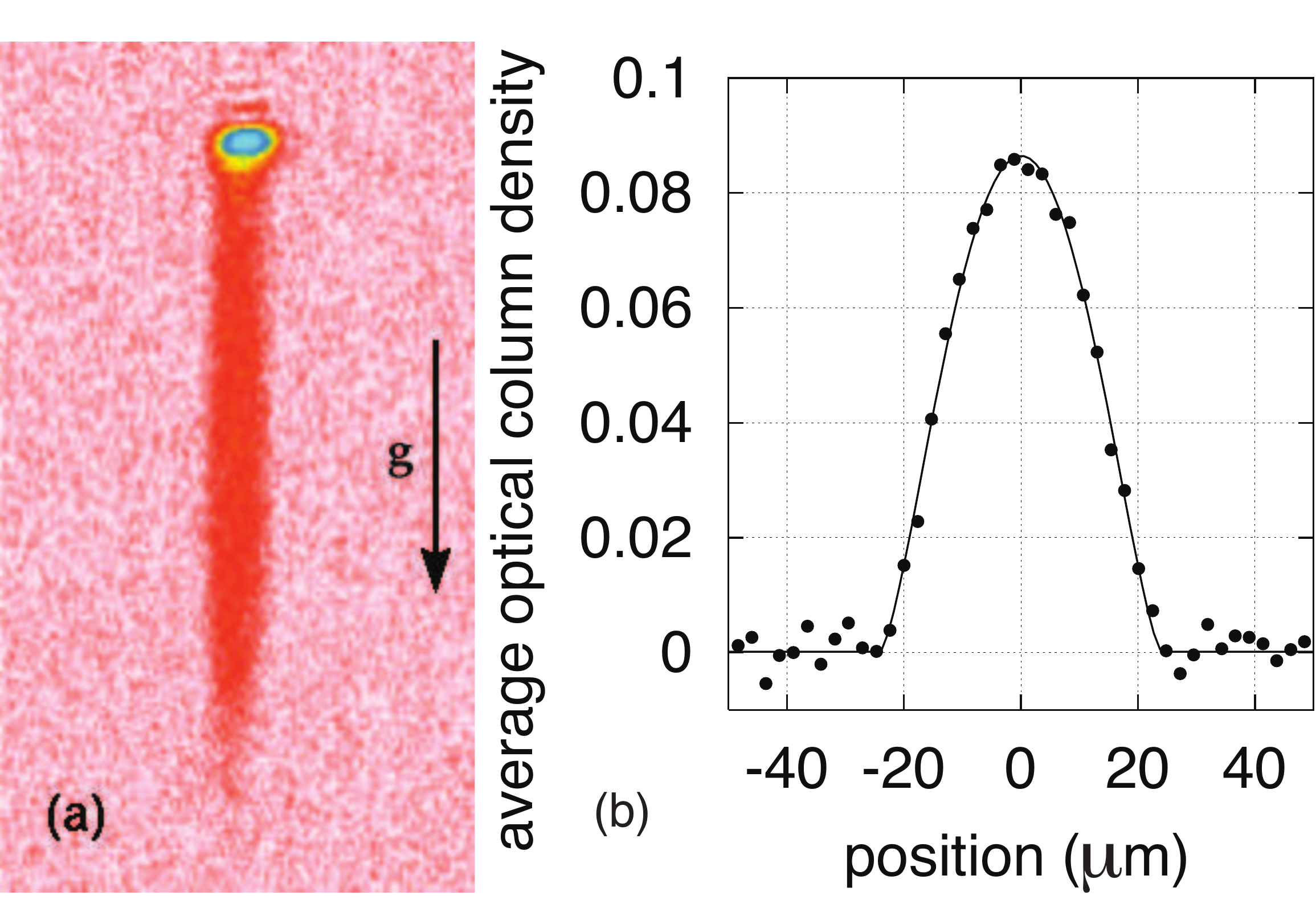}}}
\caption{(a) An atom laser beam output-coupled by spilling from an optical trap. In this case, the atoms do not experience significant mean-field repulsion whilst leaving the trap, and the momentum spread of the beam should be limited only by the uncertainty principle. The field of view comprises 0.28\,mm by 0.5\,mm. (b) Fitted transverse cut of the image averaged over a 0.19\,mm beam length. The solid line is an inverted parabolic fit to the data. Reproduced from \cite{Cennini:2003aa} with permission of the authors.}
\label{cenniniab}
\end{figure}
%%%%%

Cennini \etal\ were the first to investigate output-coupling from an optical trap \cite{Cennini:2003aa}. These authors produced condensates of 12,000 atoms distributed between the \ket{F=1,\,m_F=0,\pm1} states of \Rb{87} by evaporating atoms in a far detuned optical trap realised with a CO$_2$ laser beam. The experiment included the option of applying a magnetic field gradient toward the end of the evaporation ramp to expel atoms in field sensitive states producing a pure \ket{F=1,\,m_F=0} condensate. Atoms were output-coupled by smoothly ramping down the power in the trapping laser. The authors observed an intense beam of atoms, 1\,mm long with a transverse velocity spread less than 0.3\,mm/s (Figure \ref{cenniniab}). Although the transverse momentum spread of the beam was not measured, the authors claimed it should be limited only by the uncertainty principle, and inferred a brightness of $7\times 10^{27}$ atoms s$^2$\,m$^{-5}$ assuming a Heisenberg-limited transverse velocity distribution and a Fourier-limited axial velocity distribution.

%%%%%
\begin{figure}[t!]
\centerline{\scalebox{.7}{\includegraphics{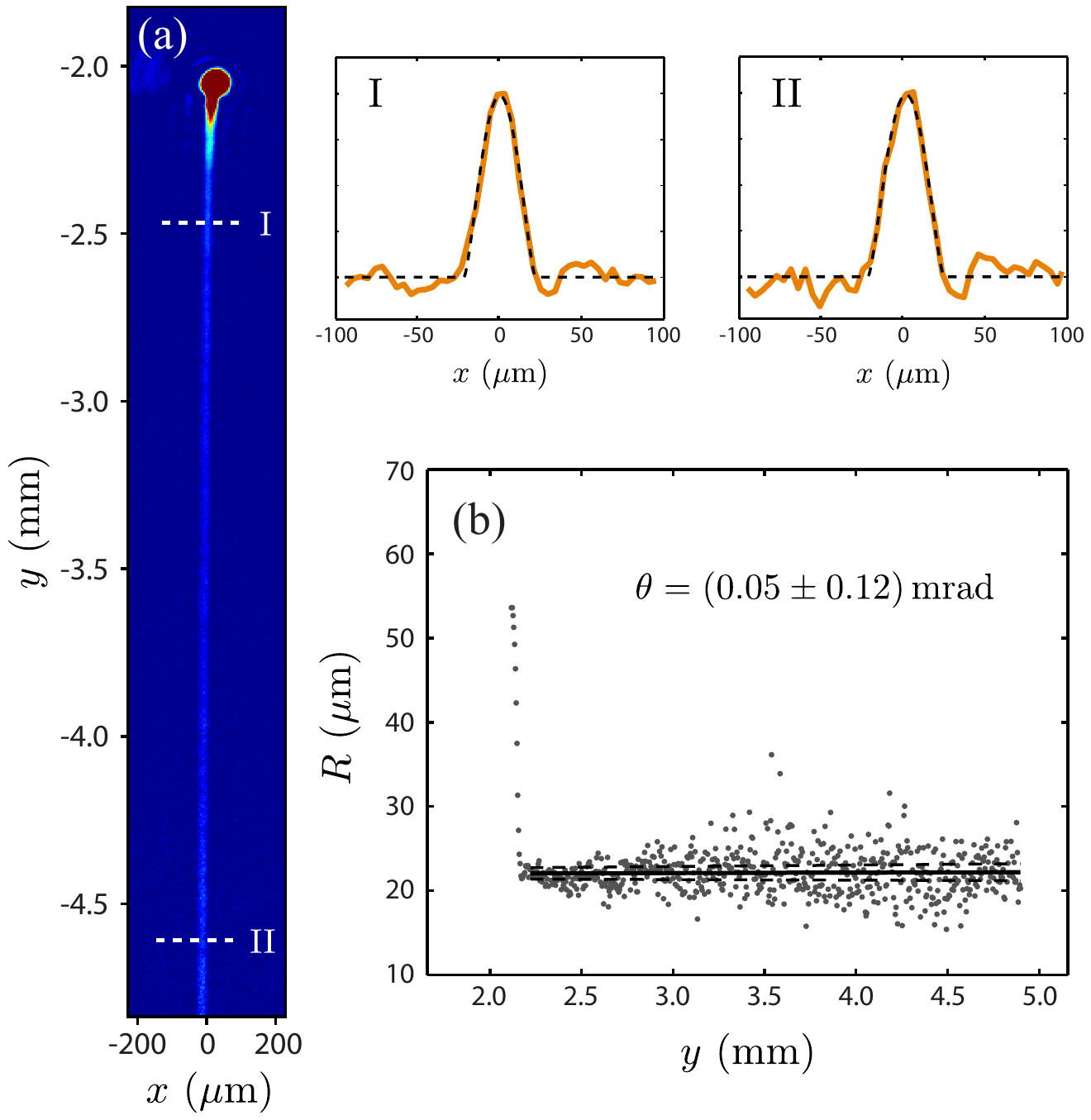}}}
\caption{A low-divergence atom laser output-coupled from a crossed optical dipole trap. (a)~Absorption image taken after 12\,ms of continuous output-coupling followed by 20\,ms of ballistic expansion. The transverse beam profile (averaged vertically over 0.1\,mm) at positions I and II are shown in the upper plots, with Thomas-Fermi fits overlaid (dashed lines). (b)~Width of the beam as a function of height, from which the divergence angle may be calculated. Each point represents a fit to a single row of pixels on the CCD image. Reproduced from \cite{Altin:2012aa} with permission of the authors.}
\label{altin-optical}
\end{figure}
%%%%%

Altin \etal\ have also investigated the spatial mode of an atom laser beam output-coupled from an optical trap by lowering the dipole laser power \cite{Altin:2012aa,Debs:2012aa}. The profile of the atom laser beam is shown in Figure \ref{altin-optical}. The transverse beam profile near each end is shown in the upper panel, with no evidence of the caustics characteristic of magnetic trap output-coupling. The divergence of the beam was determined by fitting the transverse profile to the column density expected from a Thomas-Fermi distribution: $\rho\,(1 - x^2/R^2)^{3/2}$, where $\rho$ is the peak column density. The width $R$ as a function of height is plotted in Figure \ref{altin-optical}b, and a linear fit yields a divergence angle of $\theta = (0.05 \pm 0.12)$\,mrad. This value is between 1 and 3 orders of magnitude lower than for beams produced by rf or Raman output-coupling of magnetically trapped atoms, and is within experimental uncertainty of the Heisenberg limit.

Just as the divergence of optical beams can be reduced by propagation in a single mode fibre, the divergence of atom laser beams can be reduced and the mode filtered and improved by propagation in a matter-wave guide. Guided atom lasers, in principle, allow substantially greater interrogation time in a precision inertial measurement than freely falling atom lasers and, in principle, offer the potential of improved sensitivity as a result. However, to date all precision inertial measurements with ultra cold atoms have used freely falling atoms rather than guided beams despite the reduced interrogation time. Guided atoms are strongly coupled to the environment for the entire duration of the measurement, and it is unclear at present whether guided atom lasers will be useful for precision measurements. Nonetheless, guided atom lasers are interesting devices in their own right and have been investigated by several groups.

Guerin \etal\ were the first to investigate a guided atom laser \cite{Guerin:2006aa, Bernard:2011uq}. These authors rf out-coupled a \Rb{87} Bose-Einstein condensate from a hybrid optical-magnetic trap. After forced rf evaporative cooling to 1\,$\mu$K in a pure magnetic trap, 120 mW of Nd:YAG laser light ($\lambda=1064$\,nm) was focussed to a waist of 30 $\mu$m and superimposed along the weak trapping direction of the magnetic trap. BEC was produced in the hybrid trap. The radial trap frequency ($\omega_r= 360$\,Hz) was dominated by the optical trap. Confinement in the axial direction was dominated by the weak magnetic trap with a trapping frequency $\omega_a = 35$ Hz. The $\ket{F=1,m_F=0}$ atom laser beam was rf out-coupled from the trapped $\ket{F=1,m_F=-1}$ state. To first order the $\ket{F=1,m_F=0}$ state shows no Zeeman shift, but the quadratic Zeeman shift in the direction of the guide was arranged to cancel the curvature of the optical potential producing a linear potential along the guide. Acceleration of the atoms along the guide was several orders of magnitude less than gravity, and the out-coupled atoms moved with a near-constant velocity of 9\,mm/s along the guide corresponding to a de Broglie wavelength of 0.5\,$\mu$m. The transverse mode of the laser was investigated by suddenly switching off the guide after 60\,ms of propagation. Although in theory, the experiment should have exhibited single mode propagation, experimental imperfections resulted in excitation of a small number of transverse modes.

In related work, Dall \etal\ investigated the transverse modes in a guided metastable helium atom laser out-coupled from a hybrid trap \cite{Dall:11}, and Kleine-B\"uning \etal\ investigated a gravity compensated slowly falling, guided beam by out-coupling atoms from hybrid trap and exploiting the magnetic field gradient from a quadrupole trap to balance gravity \cite{Kleine-Buning:2010kx}.

The atom laser group led by Gu\'ery-Odelin originally at \'Ecole Normale Superieure in Paris developed a scheme to produce a guided atom laser in any Zeeman state. In that experiment, $10^9$ \Rb{87} atoms were loaded from a MOT into a far detuned crossed dipole trap and evaporated by reducing the power in the optical beams \cite{Couvert:2008aa}. Evaporation could be performed with or without an applied magnetic field gradient allowing distillation of the desired Zeeman state. Out-coupling was performed by applying a magnetic field gradient along the axis of one of the trapping beams. From a measurement of the transverse velocity distribution, the authors concluded that the atoms exhibited quasi-monomode propagation, with mean excitation number in the transverse modes of $\langle n \rangle < 1$, for several millimetres along the guide. The same group published closely related theory papers on the transverse mode of guided atom lasers \cite{Gattobigio:2009vn,Vermersch:2011ys} and the coherence length of guided atom lasers \cite{Gattobigio:2010kx}.

\section{Atom laser flux} \label{sect:flux}

The sensitivity of an interferometric measurement is limited by quantum projection noise, which results from the inherently stochastic nature of the projection induced by a measurement on a quantum superposition state. In the absence of squeezing, the projection noise limited sensitivity scales with the square root of the number of particles measured. In this context, improving the flux of atom lasers is important. Bloch \etal\ were the first group to experimentally investigate the flux of an rf out-coupled atom laser \cite{Bloch:1999aa}. In their experiment, $7\times10^5$ \Rb{87} atoms were condensed in a magnetic trap with a bias field of 2.5\,G. An rf coupling field was applied for 100\,ms, and its frequency was chirped to follow the shrinking condensate. The atom laser flux was measured to be $5\times10^6$ atoms s$^{-1}$. The inferred brightness of the beam was $2\times10^{24}$ atoms s$^2$\,m$^{-5}$, many orders of magnitude greater than state-of-the-art thermal sources such as Zeeman slowers and two-dimensional magneto-optical traps.

As the out-coupling Rabi frequency was raised while holding rf frequency and pulse duration constant, the number of atoms remaining in the condensate after 20\,ms of out-coupling was found to decrease, asymptoting to a constant value set by the position of the out-coupling surface. A simple rate equation fit to experimental data indicated that loss of atoms from the trapped state due to out-coupling scaled with the square of the Rabi frequency as $\Gamma/\Omega^2= 1.2(2)\times10^{-5}$ for $\Omega \leq 20$\,kHz, where $\Gamma$ is the depopulation rate. This contrasts to the Rabi oscillations that were observed in the first pulsed out-coupling experiment performed by Mewes \etal\ and discussed earlier in this review. In a simple model that assumes the out-coupled atoms leave the out-coupling region and cannot be coupled back to the condensate, the behaviour observed by Bloch \etal\ makes sense, the stronger the rf out-coupling radiation, the higher the flux of the atom laser.

Gerbier \etal\ developed a 3D analytic theorectical description of rf out-coupling including gravity and applied it directly to the experimental results of Bloch \etal\ \cite{Gerbier:2001aa, Gerbier:2004zr}. These authors assumed the coupling was suffiiciently weak that the out-coupled atoms would not be coupled back to the condensate. Furthermore, they assumed that the only internal states that were populated in the process were the trapped $\ket{F=1,m_F=-1}$ and $\ket{F=1,m_F=0}$ untrapped states. Under these conditions, the authors developed an analytic model based on Fermi's golden rule with the trapped state being coupled to a continuum of propagating modes by the rf coupling radiation. Quantitative agreement with the experiment was obtained, and the authors emphasised the importance of gravity in the description.

\subsection{From weak to strong output-coupling: shut down of an atom laser}

In a series of experiments spanning weak and strong rf out-coupling and closely related theory, Robins \etal\ investigated classical amplitude fluctuations as a function of atom laser flux \cite{Robins:2004fk,Robins:2005aa,Dugue:2007aa}. These authors showed that, for weak rf out-coupling of both pulsed and quasi-continuous atom lasers, the flux increases as the Rabi frequency of the coupling radiation increases, in agreement with the experimental results of Bloch \etal\ and the analytic theory developed by Gerbier \etal\ However, as the Rabi frequency exceeded the weak coupling criterion, increased amplitude fluctuations and decreasing flux were observed. This represented a new strong-coupling regime, previously unexplored experimentally. Increasing the coupling strength further led to the formation of a bound dressed state that effectively shut down the atom laser. The bound state had been predicted \cite{Jeffers:2000kx, Hope:2000uq}, but until the work of Robins \etal\, it had not been observed.

Robins \etal\ made a similar series of measurements in the first experiment to investigate quasi-continuous Raman out-coupling \cite{Robins:2006aa}. Using a simple model, these authors predicted that in the weak coupling regime, the flux of Raman and rf out-coupled atom lasers would be identical at the same coupling strength. Experimentally this was found to be the case. They also predicted that the boundary between weak and strong out-coupling would be different in the two cases but were unable to verify the prediction in their early experiments due to insufficient laser power.

These authors also discussed predictions for the maximum flux of rf out-coupled atom lasers and Raman out-coupled atom lasers. Under typical current experimental conditions with condensates of $10^6$ atoms at a density of $10^{13}$ cm$^{-3}$, the weak/strong out-coupling boundary was predicted to give a peak flux of $1.4\times10^8$ atoms/s for rf atom lasers and $4.2\times10^9$ atoms/s for Raman atom lasers. The reasoning behind this prediction was initially based on a classical description of the centre-of-mass motion but a quantum description of internal degrees of freedom. In such a model, weak out-coupling occurs when the Rabi period is much longer than the time for atoms to leave the out-coupling region. The Born-Markov approximation, first discussed in the context of atom lasers by Moys \etal, applies in this regime \cite{Moy:1999fk}. In the strong out-coupling regime, the Born-Markov approximation does not apply; the Rabi period is shorter than the timescale for atoms to leave the out-coupling region and they are back-coupled from the beam to the trapped state. This disturbs the condensate, produces fluctuations in the output beam and eventually shuts down the atom laser due to the formation of a bound rf dressed state.

\subsection{Flux boosting from Raman output-coupling}

Robins \etal\ argued that the momentum kick provided by Raman out-coupling would remove atoms from the out-coupling region more quickly, that the boundary between weak and strong out-coupling would be higher for a Raman atom laser than an rf atom laser, and that the maximum flux of Raman atom lasers would thus be correspondingly higher. Dugu\'e \emph{et al.} further showed theoretically that a two-state out-coupler (one involving only a single trapped and untrapped state, with no other couplings) would produce maximum flux \cite{Dugue:2007aa}.

In 2010, Debs \etal\ made the first detailed experimental comparison of Raman and rf atom lasers spanning the strong and weak out-coupling regimes, which verified predictions that a two-state Raman out-coupler will produce the maximum flux atom laser from magnetically confined samples \cite{Debs:2010aa,Debs:2009aa}. In addition, these authors described weak/strong out-coupling in terms of adiabatic/diabatic following at an avoided crossing of the dressed state potentials. This model also predicts that the weak/strong out-coupling boundary occurs at higher out-coupling strength for Raman atom lasers in comparison with rf atom lasers. 

Bernard \etal\ made a detailed experimental and theoretical study of the flux of an atom laser rf out-coupled into a horizontal guide \cite{Bernard:2011uq}. In this case, the only force removing atoms from the out-coupling region is the mean field repulsion from the atoms in the condensate. Consistent with the work of Robins \etal\ on flux limits of freely falling beams, these authors concluded that the maximum flux of a horizontally guided atom laser is substantially less than other schemes due to the weak extraction from the out-coupling region. The authors discuss the applicability of the scheme for the study of quantum transport phenomena.

\subsection{Methods for further increasing flux}

The current bottleneck in raising the average flux of atom lasers is the rate at which cold atoms can be condensed to produce the source Bose-Einstein condensate. The highest flux BEC apparatus that we are aware of produces $2.4\times10^6$ atoms per second \cite{Stam:2007aa}, 25 times lower than the best reported thermal source used in a precision inertial measurement \cite{Muller:2008ab}. In the final paragraphs of this section, we briefly discuss progress on alternative high flux sources for atom lasers.

After a proposal by Manndonet \etal\ \cite{E.-Mandonnet:2000fk}, the group of Gu\'ery-Odelin pursued a high flux atom laser by the evaporation of atoms in a long guide \cite{E.-Mandonnet:2000fk, P.-Cren:2002oq,Lahaye:2004ys,T.-Lahaye:2005tg,Lahaye:2006ly,Lahaye:2006qf,Reinaudi:2006zr}. The idea was to inject non-degenerate atoms from a MOT into the guide and to evaporate as the atoms move along the guide. The technique effectively parallelises the steps required to cool to degeneracy by replacing the traditional temporally separated sequence of laser- and evaporative-cooling steps with a spatially separated one. In theory, atoms are injected at one end of the guide at low phase space density and exit the other end as a degenerate atom laser beam. The guide in these experiments was formed from four parallel 4.5 metre long copper tubes. The tubes were water cooled and carried a current in excess of 300\,A to produce a two-dimensional quadrupole field. Evaporation along the guide was performed via a series of rf antennas or by removal with a material barrier. The collisional regime was reached, but to date it has proved difficult to produce large gains in phase space density with this approach. The idea is very appealing, but the apparatus in its present form is large, and if this approach were successful in the future, it would appear to be best suited to laboratory applications rather than in field-deployable precision measurement devices. Complementary to the work by Gu\'ery-Odelin, Raithel's group at the University of Michigan has also studied continuous cooling in waveguides, and recently developed a more compact spiral geometry for the guide \cite{Teo:2001vn,Olson:2006kx,Power:2012}.

The Harvard University group led by Doyle has pioneered the use of conventional cryogenics to buffer-gas-cool molecules and alkali gases to several Kelvin (see for example Refs.\ \cite{Magkiriadou:2011vn,Hong:2009cr, PhysRevLett.95.173201,Egorov:2002dq,R.-deCarvalho:1999ij} and references therein). This would need to be followed by evaporative cooling or some other process to cool to degeneracy. In the context of high flux atom lasers, the technology is interesting to watch, but at present, it is unclear how successful this approach will be as a first stage cooling process along the path to high flux degenerate samples. 

The use of inhomogeneous pulsed electric or magnetic fields to slow atomic and molecular beams has been investigated by a number of groups. The idea is straightforward: a pulsed beam of alkali atoms in a low field seeking state approaches a current-carrying coil along its axis. The atom pulse climbs the potential hill produced by the field at the expense of kinetic energy. As the pulse reaches the top of the hill, the magnetic coil is turned off. A series of coils of this kind fired in a timed sequence can be used to slow a beam of alkali atoms. A 64 stage atomic coil-gun of this kind was developed and investigated by Narevicius \etal\ \cite{Narevicius:2008kx}. These authors slowed an atomic beam of metastable neon from 447\,m/s to 56\,m/s, removing 98\% of the kinetic energy of the beam. Similar approaches using Zeeman decelerators for atoms and Stark decelerators for polar molecules have been pioneered in the group of Meijer in Berlin and by others (see Ref.\ \cite{Meerakker:2008bs} and references therein). Again, it is unclear whether this technology can be successfully employed as a first cooling stage in high flux atom lasers, but it is certainly an interesting and promising technology.

In an ongoing effort to realise a continuous atom laser \cite{Schmidt:2003,Aghajani:2009}, the team of T. Pfau recently demonstrated continuous loading of a conservative trap using a one-way atomic diode \cite{Falkenau:2011uq,Falkenau:2012}. In their experiment, a beam of atoms in a low field seeking state sourced from a moving molassses MOT is guided to a hybrid optical-magnetic trap. The optical component of the trap is produced by focussing 80\,W of 1070\,nm light to a waist of 30\,$\mu$m. Magnetic coils are arranged to produce a potential hill equal in height to the kinetic energy of the atoms. As the atoms move along the guide in the low field seeking state, they climb the potential hill converting kinetic energy to potential energy. At the top of the hill, the atoms encounter an optical pumping beam that pumps them to the maximally stretched high field seeking state that is trapped in the combined optical and magnetic fields. The high field seeking state is dark to the optical pumping beam and no longer interacts with the optical pumping laser, while the potential energy is dissipated as spontaneous emission in the process. The spontaneously emitted photons leave the trapping region ensuring the process is irreversible at least for samples that are not too optically dense. The hybrid trapping potential has a depth of 5\,mK axially (along the optical trapping beam) and 1\,mK in the radial direction. The setup achieves a loading rate of $2\times10^7$ atoms s$^{-1}$ with a peak density of $4\times10^{11}$ atoms cm$^{-3}$. As the trap loads, the density and collision rate build. While still loading, atoms begin to evaporate from the trap, increasing the phase space density by roughly two orders of magnitude. As the authors point out, this technology could be combined with an intense atomic or molecular beam that is initially slowed using Zeeman or Stark deceleration as described in the previous paragraph. Again, this is a very promising technology that may be applicable to high flux, possibly continuous atom lasers.

\section{Coherence} \label{sect:coherence}

In this section, we concentrate on the coherence properties of Bose-Einstein condensates and atom lasers \cite{Kasevich:2002kl}. It is the high spectral density and brightness that results from the high first order coherence of unfiltered laser sources that are often the most useful properties of lasers compared with thermal sources. It is, however, the higher order coherences that strictly distinguish laser sources from thermal sources. We review the progress of experimental work which has led to the appreciation of atom lasers as truly analogous to optical lasers, well-described by a state exhibiting coherence to all orders.

\subsection{First-order coherence}

The MIT group were the first to experimentally investigate coherence and long range correlations in Bose-Einstein condensates \cite{Andrews:1997aa, Ketterle:1997oq}. In their experiment, two pure BECs well below the phase transition were produced in a double well potential formed from a magnetic trap divided along its centre by a blue-detuned light sheet. The magnetic trap and light sheet were simultaneously extinguished. The expanding condensates overlapped and were imaged using absorption imaging showing high contrast fringes -- demonstrating high first order coherence from an unfiltered source. This result was an important step in the characterisation of BECs and atom lasers. In the same paper, the authors allude to an experiment that also demonstrated first-order coherence of a pulsed atom laser based on rf output-coupling. 

In a later paper, the MIT group used Doppler-sensitive Bragg spectroscopy to measure the momentum distribution of both trapped and freely falling condensates \cite{Stenger:1999a}. The momentum distribution in-trap is broadened by the finite size of the condensate as determined by mean-field repulsion and the trapping frequencies. In measurements performed in-trap, the coherence length of the condensate was shown to be equal to its size. On release, mean field energy is converted into kinetic energy, thus broadening the momentum distribution. 

Contemporaneously, the NIST group performed a series of interferometric experiments also studying the momentum width and first-order coherence of trapped and freely falling condensates. They found a uniform phase in trapped BECs and a large spatial variation in phase in freely falling condensates, consistent with the mean field expansion of the freely falling atoms \cite{Hagley:1999zr,Simsarian:2000aa}. In a related experiment, Bloch \etal\ studied the first order coherence of condensates and thermal clouds close to the BEC phase transition by measuring the fringe visibility produced via interference of two atom laser beams sourced from spatially separated regions of the condensate, finding a dramatic increase in fringe visibility as the temperature of the atomic sample passed through the BEC transition. The result is consistent with the sudden decrease in momentum width and consequential increase in first order coherence that characterises the BEC transition \cite{Bloch:2000ly}.

K\"ohl \etal\ measured the temporal coherence of an rf atom laser beam by reflecting a matter-wave from a potential barrier and exploiting magnetic resonance imaging to achieve high spatial resolution (65 nm) fringe contrast measurement \cite{Kohl:2001vn}. They found that the momentum width of the out-coupled pulse was Fourier limited by the duration of the output-coupling pulse, a finding borne out in a later theory paper by Johnsson \etal\ \cite{Johnsson:2007aa}.

All of these early experiments demonstrated first-order coherence, a property that depends on the momentum distribution of the source and is independent of other details of the many-body state. Two interfering Fock states, a Poissonian mixture of Fock states, two coherent states with definite or indefintite phase or even two thermal states all with the same momentum width (and therefore the same first-order coherence) would exhibit interference fringes with the same fringe visibility in any single realisation of an experiment. 

The mechanism of the establishment of a relative phase between two interfering cold atom sources has been discussed in an interesting series of papers on the subject \cite{Javanainen:1996vn,Cirac:1996uq,Naraschewski:1996kx}. The aspects of fringe contrast, long range order, coherence and number squeezing in a system described by the Bose-Hubbard Hamiltonian was investigated experimentally by Orzel \etal\ \cite{Orzel:2001aa} and by Greiner \etal\ \cite{Greiner:2002aa, Greiner:2002fk}, who studied the superfluid to Mott insulator transition for BECs trapped in an optical lattice. When the tunnelling term in the Bose-Hubbard Hamiltonian dominated at low lattice depth, high visibility fringes were observed. As the lattice depth was increased, the ratio of the on-site interaction term to the tunnelling term increased, producing number squeezed states at each lattice site and causing the visibilty of interference fringes to decrease. The loss of fringe visibility observed in the insulating (deep lattice) regime was interpreted as a loss of long range phase coherence. This interpretation was discussed in some depth in a paper by Hadzibabic \etal\ who performed a similar experiment in a one dimensional lattice \cite{Hadzibabic:2004uq}. These papers forms an interesting literature on the interpretation of fringe contrast measurements with BECs, and we refer the reader to them for more detailed discussion.

\subsection{Higher-order coherence}

Higher-order correlation functions, expressing the conditional probability of measuring an atom at a particular position or time, are markedly different for condensates and thermal clouds, as is the case for thermal and coherent light sources in optics. A measurement of higher (than first) order coherence clearly distinguishes between a thermal source and a laser source. Following an idea by Kagan and Shlyapnikov \cite{Kagan:1985aa}, Burt \etal\ used three-body recombination to compare density fluctuations in Bose-Einstein condensates and cold thermal gases, finding that the three-body recombination rate was a factor of $3! = 6$ higher than in a thermal cloud, in agreement with the theoretical prediction. This experiment clearly demonstrated the third-order laser-like coherence property of trapped condensates \cite{PhysRevLett.79.337}.

The first experiment probing higher-order coherence in a freely propagating atom laser beam was performed by the group at ETH. In their experiments, atom detection by means of an optical cavity was employed to investigate the first- and second-order temporal coherence of an atom laser beam and the spatial coherence of Bose-Einstein condensates \cite{Ottl:2006aa,Bourdel:2006aa}. An optical Fabry-P\'erot cavity with a finesse of $3\times10^5$ operating in the strong-coupling regime of cavity QED was located 36\,mm below the condensate. The transmission of the probe beam through the cavity was measured with a single photon counting module. The presence of an atom in the cavity led to either an increase or decrease of the transmitted field depending on the detuning of the probe laser from cavity resonance.

\"Ottl \etal\ used this setup to measure the second-order temporal correlation function of an rf out-coupled atom laser beam, observing a constant second order correlation correlation function $g_2(\tau) =1.00\pm0.01$ \cite{Ottl:2005aa}. This was the first result to measure a corrlation function that clearly distinguished freely propagating atom lasers from thermal atomic sources. For lasers $g_2(\tau=0) =1$, while for thermal sources $g_2(\tau=0) = 2$ \cite{Yasuda:1996fk}. In the same apparatus, \"Ottl \etal\ measured $g_2(\tau=0)$ close to 2 for a pseudo-thermal source. In a beautiful experiment, the ETH group used cavity detection to study the formation of long-range order in real time during Bose-Einstein condensation by recording the fringe visibility of interference between two atom laser beams out-coupled from different regions of the condensate following a sudden quench \cite{Ritter:2007ys}. This work was closely related to the earlier work by Bloch \etal\ on the coherence properties of thermal clouds and BECs close to the BEC phase transition \cite{Bloch:2000ly}.

%%%%%
\begin{figure}[t!]
\centerline{\scalebox{1}{\includegraphics{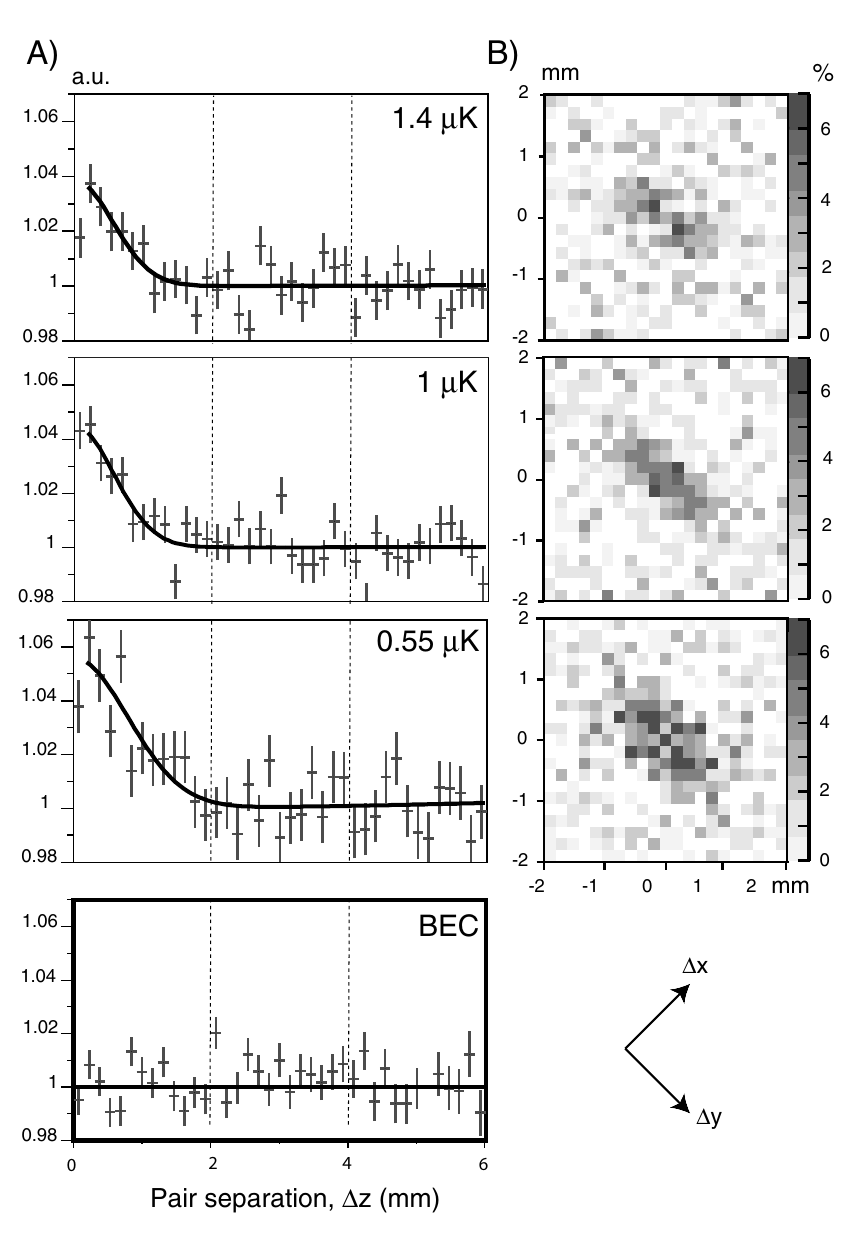}}}
\caption{Measurements of the second-order correlation function for atomic clouds above and below the BEC transition temperature. (a) Normailzed correlation functions along the vertical ($z$) axis for thermal gases at three different temperatures and for BEC. For the thermal clouds, each plot corresponds to the average of a large number of clouds at the same temprature. Error bars correspond to the square root of the number of pairs. a.u. -- arbitrary units. (b) Normalised correlation function in the $\Delta x-\Delta y$ plane for the three thermal gas runs. The arrows at the bottom show the $45^\circ$ rotation of the coordinate system with respect to the axes of the detector. Reproduced from \cite{Schellekens:2005ys} with permission of the authors.}
\label{schellekens}
\end{figure}
%%%%%

Schellekens \etal\ studied two-body correlations in freely falling metastable helium clouds both above and below the phase transition \cite{Schellekens:2005ys}. These authors clearly observed bunching of atoms in second-order correlation measurements on thermal clouds above the BEC transition temperature and the absence of bunching in measurements made on condensed clouds (Figure \ref{schellekens}). While the work of 
\"Ottl \etal\ was a Hanbury-Brown-Twiss (HBT) experiment in one dimension, the work of Schellekens \etal\ was a HBT experiment in all three dimensions achieved by exploiting the high temporal and spatial resolution provided by the high internal energy of metastable helium, a multi-channel plate and a delay line detector. The measurements are analogous to correlation measurements in optics that allow thermal light sources to be distinguished from optical lasers.

Jeltes \etal\ compared second-order correlations for a low density Bose gas (metastable $^4$He) and a low density Fermi gas (metastable $^3$He) under similar conditions in a HBT-type experiment \cite{Jeltes:2007uq}. The density was kept low so that atomic interactions were negligible. They observed bunching for thermal bosons and anti-bunching for fermions in agreement with theory. In closely related work and again using similar technology, Manning \etal\ observed the HBT effect using pulses rf out-coupled from a metstable helium BEC \cite{Manning:10}. Dall \etal\ observed matter-wave speckle in the density distribution of guided matter-waves and discussed the relationship of speckle to the HBT effect \cite{Dall:2011fk}. Hodgman \etal\ have measured third- and higher-order coherences for an atom laser coupled into a waveguide \cite{Hodgman25022011,Manning:pc}.

Although higher-order coherence is not critical for precision measurements of the type discussed in this review, it is important to realise that the measurements of atom laser coherence to high orders formally establishes the atom laser as truly analogous to an optical laser, in the sense that its quantum state is well approximated by a coherent state.

\section{Squeezing} \label{sect:squeezing}

Squeezing a coherent state such as a Bose-Einstein condensate results in reduced fluctuations in some observable (usually in the amplitude or phase quadrature) at the expense of increased fluctuations or variance in the conjugate variable. Squeezed states are of great fundamental interest as they can be used to produce entanglement and study nonlocality in quantum mechanics, but they can also be used to perform measurements with reduced quantum noise. Rigorous theoretical treatments of this subject from the point of view of optics can be found in the texts by Gerry and Knight \cite{Gerry:2004aa} and Walls and Milburn \cite{Walls:2008aa}.

With careful design and implementation, it is possible to exploit squeezing to create a measurement device with sensitivity exceeding the limit imposed by quantum projection noise \cite{THORNE:1978ly,CAVES:1980zr,CAVES:1981ve}. However, this is not necessarily the easiest way to increase sensitivity. In the field of optics, it has nearly always proven possible and preferable to increase the flux of the source rather than to squeeze in order to enhance the signal-to-noise ratio of a measurement. Atoms are different. Cold atoms are relatively expensive to produce and to date it has proven difficult to significantly increase the flux of atomic sources. Squeezing in this circumstance appears to be a viable way forward, especially considering that high quantum efficiency detection has already been developed by several groups \cite{Ottl:2006aa,Teper:2006aa,Heine:2010aa,Bakr:2010aa,Sherson:2010aa,KohnenM.:2011aa,poldy:013640,Takamizawa:2006aa,Poldy:2012}, and that atom interferometers operate in the low loss environment of ultra-high vacuum.

Much has been written about measurement sensitivity that exceeds the quantum projection noise limit, Heisenberg and super-Heisenberg scaling of the sensitivity of measurements and entanglement in atomic systems \cite{Kitagawa:1993aa,Holland:1993zr,Bouyer:1997fk,Dowling:1998fk,Pu:2000ly,Duan:2000uq,Sorensen:2001vn}. These considerations are both fascinating and important not only to future metrology but also in fields such as quantum computing, quantum information, quantum emulation and fundamental physics. However, in practical metrology, where the goal is to measure a quantity such as time, gravity, acceleration, rotation or a magnetic or electric field, what matters are things such as sensitivity, bandwidth, Allan deviation, duty cycle, mass and volume of the apparatus, immunity to vibration and a host of other practical considerations. It is unimportant whether the beams in the measuring interferometer are entangled or not, whether the source is squeezed or classical or how the sensitivity scales with flux.

For metrology, if squeezing allows us to access a useful parameter regime in a precision measurement that would be inaccessible or more difficult, bulky or costly by other means, then it should be applied -- otherwise, it clearly should not. As we discuss in this review, squeezing atomic systems may be a viable route to improve the sensitivity of precision inertial measurements, and high-flux squeezed atomic sources and atom lasers would undoubtedly find uses in metrology.

The realisation of a freely propagating squeezed atom laser and its application to precision measurement is an outstanding goal of the atom optics community. In this section, we will be guided in our discussion by the current experimental literature, and refer to only a part of the extensive theoretical literature on the subject. It should be noted that spin squeezing via interaction with a light field has been intensively studied in thermal samples (see for example \cite{Appel:2009aa,Chen:2011uq,Leroux:2010fk}). The work in thermal samples is relevant to squeezing in atom lasers but outside the scope of this review. The interested reader should consult the excellent review article by Hammerer \etal\ and the references therein \cite{Hammerer:2010ys}. 

Statistics, noise and quantum squeezing are interesting and important quantities in both photonic and atomic many-body fields. The sensitivity of light-atom precision interferometric measurements is limited fundamentally by the quantum noise that characterises the atomic source. Although squeezing the quantum noise on an atomic source is difficult and must be combined with low losses and high quantum efficiency detection if it is to be useful for metrology, there have been some exciting advances in recent years, particularly for small samples of atoms. The outstanding challenge is to squeeze a large number of atoms and obtain a result that is significant to metrology. Squeezing optical sources has yielded a vast number of fundamental results and remains central to experimental quantum optics, a field that has been intensively researched for more than 20 years \cite{Bachor:2004}. Recently, squeezed light was successfully used to improve the sensitivity of the GEO600 gravitational wave detector \cite{Collaboration:2011aa}.

In what follows, we highlight the different techniques that have been investigated for producing squeezed atomic states in Bose-condensed system. With the exception of quantum state transfer, all of these processes rely on the nonlinearities present in atomic systems due to atomic interactions. Quantum non-demolition measurements \cite{Kuzmich:2000,Julsgaard:2001,Schleier-Smith:2010} can also be used to generate squeezing, but these techniques are not specific to BECs or atom lasers and will not be discussed here.

\subsection{Quantum state transfer}

Transfer of a squeezed state from an optical beam to a condensate or atom laser has been considered by several groups \cite{Jing:2000fk, Fleischhauer:2002uq, Haine:2005aa, Haine:2006aa}. For example, Haine \etal\ discuss Raman out-coupling an atom laser beam in a scheme where one of the optical Raman beams is squeezed. The basic idea is fairly intuitive: if the photons in the Raman beam are anti-bunched (squeezed), then since each out-coupled atom must absorb from or emit into the optical beam, under appropriate conditions the stream of atoms falling out of the condensate will also exhibit anti-bunching. Thus, optical squeezing can be transferred to the atom laser, and the Raman beam and the atomic beam are entangled after the interaction. This entanglement of photonic and atomic beams is interesting and potentially useful not only in metrology but also for quantum information technology \cite{Olsen:2008vn}. However, to produce a squeezed atom laser beam, such schemes are experimentally complex, and the degree of optical squeezing available is not particularly high. For these reasons, quantum state transfer schemes have not yet been pursued experimentally.

\subsection{Squeezing via state-changing collisions}

Several methods have been investigated for producing squeezing via inelastic atomic collisions. These can be considered analogous to optical parametric down-conversion. Conservation of angular momentum in spin-exchange collisions between two atoms in a well-defined Zeeman state \ket{m_F} can result in an entangled state with one atom in the \ket{m_F+1} and the other in the \ket{m_F-1} state. This situation was first examined theoretically by Duan \etal\ \cite{Duan:2000uq} and Pu and Meystre \cite{Pu:2000ly}. In 2011, the group of C. Klempt realised these proposals experimentally, using spin exchange collisions between \Rb{87} atoms in the \ket{F=1,\,m_F=0} state to populate the \ket{m_F = \pm1} states. The total number of atoms produced in the $\ket{m_F = \pm1}$ states and the sum of the populations increased exponentially in time. The conjugate variable, the sum of the phases, was exponentially damped in time. With a total number of 7800 atoms, spin noise supression of 6.9\,dB below the shot noise limit was observed, and the authors performed an interferometric experiment concluding that the twin beams were entangled and could be applied to sub-shot-noise-limited interferometry.

Counter-intuitively, given the deleterious effect of losses on squeezing in optics, inelastic processes which cause loss of atoms from a sample can also generate squeezing. The rate of three-body recombination, for example, depends on the probability of finding three atoms close together, and is thus higher in regions of higher density. Three-body loss can therefore reduces density fluctuations in a sample, effectively anti-bunching the atomic state. Itah \etal\ \cite{PhysRevLett.104.113001} and Whitlock \etal\ \cite{Whitlock:2010uq} have used three-body recombination to produce sub-Poissonian variances in the number of atoms per site in an optical lattice.

In 2004, it was suggested that a BEC of composite molecules could be dissociated to produce a pair of entangled atom laser beams in a realisation of an Einstein-Podolsky-Rosen scenario \cite{Drummond:2004kx}, but this proposal has yet to be studied experimentally.

B\'ucker \etal\ have observed analog parametric down-conversion via collisional de-excitation of a one-dimensional BEC in a waveguide, producing two entangled, guided atom laser beams \cite{Bucker:2011kx}. In this experiment, a quasi-condensate of 700 atoms was produced in the ground state of a hybrid magnetic/optical trap with weak confinement along the axial (waveguide) direction. An excited transverse mode of the waveguide was populated by shaking the trap along an optimised trajectory, producing almost complete population inversion. Binary collisions between atoms in the excited mode then repopulated the ground transverse mode, and conservation of momentum required that the atoms ejected into the ground state transverse mode were emitted in opposite directions along the guide. Bose-stimulation of the process led to exponential growth of the population of the ground state transverse mode. The waveguide geometry of this experiment produced guided, entangled beams with a measured variance in the number difference approximately ten times below that expected for Poissonian statistics.

%%%%%
\begin{figure}[t!]
\centerline{\scalebox{1}{\includegraphics{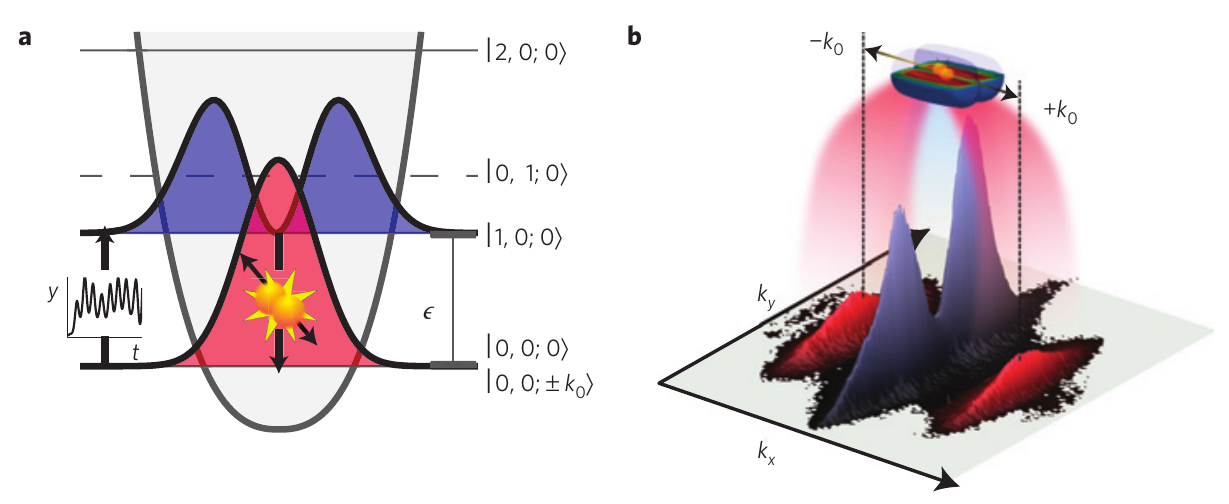}}}
\caption{Production of entangled, guided atom laser beams. (a) A one-dimensional quasi-BEC is transferred from the ground state \ket{0,0;0} into \ket{1,0;0}, the first excited state of the trapping potential along the radial $y$-direction. This is accomplished by means of a fast non-adiabatic movement of the potential minimum along an optimized trajectory. The excited state decays by emission of twin-atoms into the radial ground state modes \ket{0,0;\pm k_0} (b) After excitation and pair emission, the cloud is released from the trapping potential and imaged after expansion. The central part of the system clearly shows the spatial structure of the radially excited state (blue). Two clouds containing the twin-atoms (red) are emitted. Reproduced from \cite{Bucker:2011kx} with permission of the authors.}
\label{fig:twinatombeams}
\end{figure}
%%%%%

\subsection{Four-wave mixing}

Atomic four-wave mixing (4WM) has been studied intensively as a means to produced entanglement in BECs and atom laser beams \cite{Lenz:1993fk,Goldstein:1995uq}. Deng \etal\ were the firt to observe 4WM in a BEC \cite{Deng:1999cr}. In their experiment, Bragg diffraction of an expanded sodium condensate was used to create the three input matter-wave beams that are necessary in a 4WM experiment. Two components of the condensate were interfered to produce a matter-wave grating, of which a third was diffracted. In addition to the creation of a fourth beam by 4WM, the input matter wave was amplified by a factor of 1.5 by Bose stimulation.

Vogels \etal\ observed strong amplification of the seed beam in a similar 4WM experiment also performed with a sodium BEC \cite{Vogels:2002ys,Vogels:2003vn}. The observed amplification of the weak input beam by a factor of 20 suggested that the variance of the number difference of the amplified input beam and of the fourth beam created in the 4WM process would be suppressed by a factor of $\sqrt{40}$.

RuGway \etal\ measured and compared the second-order correlation function for spontaneous and amplified 4WM with an atom laser out-coupled from a metastable helium BEC \cite{RuGway:2011kx}. In this experiment, the rf out-coupling surface was placed in a high density region of the condensate and operated in the strong out-coupling limit. The high aspect ratio BEC used in this experiment resulted in amplification along the long axis of the condensate, and an atom laser profile with four peaks arising from amplified 4WM was observed. In the early stages, spontaneous 4WM dominated, producing an $s$-wave shell of scattered atoms. As the matter-wave grating formed, specific modes were amplified leading to the observed four peaked atom laser profile. The second-order correlation function was calculated both for the spontaneous process that produced the $s$-wave shell and the amplified process. The spontaneous process exhibited bunching, the amplified process did not. In related experiments, Pertot \etal\ experimentally realised collinear 4WM \cite{Pertot:2010kx}.

Perrin \etal\ were the first to report the observation of correlations and and number difference squeezing in a 4WM experiment \cite{Perrin:2007ly,Perrin:2008kx, Krachmalnicoff:2010zr,Jaskula:2010uq}. In this experiment, two metastable helium BECs were collided and the atoms scattered in the collision process recorded using a multichannel plate/delay line anode single-atom counter. This detection scheme affords high temporal and spatial resolution, and is uniquely applicable to metastable atoms. In contrast to earlier experiments, this experiment was unseeded and relied on spontaneous 4WM. Clear correlations were observed between atoms with opposite momenta, as well as Hanbury Brown-Twiss correlations for collinear momenta.

\subsection{Squeezing via condensate self-interaction}

Fortuitously, it happens that the correlations induced by elastic $s$-wave scattering in a condensate can also be harnessed to generate squeezing and entanglement. This is because the Hamiltonian that represents the two-body interaction potential is identical in form to that describing the Kerr effect in a nonlinear crystal. Atomic interactions are also important in the formation of a Bose-Einstein condensate, and most BECs are created with a large positive scattering length, corresponding to repulsive atom-atom interactions. This fact was exploited in the first experimental demonstration of squeezed atomic states in a Bose-Hubbard system realised by loading a BEC into an optical lattice \cite{Orzel:2001aa,Greiner:2002aa, Greiner:2002fk} (these experiments were discussed in Section \ref{sect:coherence}). When the repulsive interactions are large compared with the tunnel coupling between neighboring lattice sites, it is energetically favourable for have the same number of atoms in each well, and sub-Poissonian relative number fluctuations can occur. In these experiments, the reduced number uncertainty could not be measured directly, but was instead inferred from interference patterns obtained when the atoms were released from the lattice.

Squeezing via self-interaction in a BEC was first directly observed in an experiment by Est\`eve \etal\ in 2008 \cite{Esteve:2008aa}. \Rb{87} atoms were distributed over a small number (two to six) of sites in a one-dimensional optical lattice, with $100-1000$ atoms per site. Reduced fluctuations in the atom number difference between two adjacent wells were directly measured using high-resolution absorption imaging, showing relative number squeezing of over 7\,dB.

Absolute number squeezing in small trapped samples was first observed by Chuu \etal\ \cite{Chuu:2005kx}. In this work, an ultrastable optical dipole trap was used to confine a degenerate Bose gas of a few hundred atoms. Due to interactions, the atom number is related to the trap depth (as $N\propto U_0^{5/2}$ in the Thomas-Fermi limit). By precisely controlling the depth of the trap, these authors were able to control the number of atoms held, measuring fluctuations in atom number of 40\% below the expected Poissonian fluctuation for a mean number of 60 atoms.

The most studied method for producing squeezing via self-interaction in Bose-condensed systems is the the one-axis twisting scheme, first proposed by Kitagawa and Ueda in 1993 \cite{Kitagawa:1993}. This method exploits a difference in the interaction strength between atoms in two different states to produce so-called ``spin-squeezed states'', states with reduced uncertainty in the fictitious spin projection $J_z$ (equivalent to the $z$-projection of a state vector on the Bloch sphere) or in an orthogonal observable \cite{Sorensen:2001}. An initial coherent state is placed into an equal superposition of states \ket{1} and \ket{2}, with equal uncertainty in $J_x$ and $J_y$ (or in relative number and phase). Under the influence of the Kerr nonlinear Hamiltonian $\hat{H}\sim J_z^2$, the symmetric uncertainty of this state undergoes a shearing transformation, evolving to an asymmetric uncertainty with reduced variance in one direction -- a spin-squeezed state. A rigorous theoretical description of spin squeezing may be found in Refs.\ \cite{Kitagawa:1993,Li:2009,Ma:2011}. Spin squeezing in two-state systems is of particular interest in the context of interferometry, as discussed in the following section.

There has been extensive theoretical work done on spin squeezing via one-axis twisting in Bose-Einstein condensates (see \S3.1 of Ref.\ \cite{Ma:2011} and references therein). Considerably fewer authors have investigated specifically the squeezing of an atom laser. Jing \etal\ considered quadrature squeezing of an atomic beam via atom-atom interaction using a zero-dimensional single-mode model \cite{PhysRevA.65.015601}, but little squeezing was found. In 2007, Johnsson and Haine studied quadrature squeezing in a simulation of a Raman out-coupled \Rb{87} atom laser modelled using stochastic phase space methods. The stochastic model predicted squeezing of up to 10\,dB in reasonable agreement with a single-mode analytic model \cite{Johnsson:2007ac}. Although the scheme appears to be comparatively simple, the detection of quadrature squeezing requires a strong classical local oscillator. For photons this is straightforward, but an interacting atomic local oscillator will self-squeeze. This represents a significant obstacle to observing quadrature squeezing in atomic systems.

\subsection{Spin squeezing and interferometry}

In an interferometric measurement, the relative phase acquired by two components of a coherent superposition is determined by conversion into an amplitude difference which can then be detected. The inherent phase uncertainty of a coherent state limits the precision that can be attained in such a measurement. The phase sensitivity $\Delta\phi$ achievable in an interferometer whose source is an $N$-particle coherent state (such as an unsqueezed BEC or atom laser) is equivalent to that obtained with $N$ independent particles or $N$ measurements of a single particle: $\Delta\phi = 1/\sqrt{N}$. States with reduced phase uncertainty -- spin-squeezed states -- allow the sensitivity to exceed this value. The amount of squeezing can be quantified by the coherent spin squeezing factor $\xi_S$, which gives the minimum interferometric phase uncertainty $\Delta\phi = \xi_S/\sqrt{N}$. A value of $\xi_S < 1$ corresponds to metrologically useful squeezing. A detailed review of quantum spin squeezing and its application to interferometry is given in Ref.\ \cite{Ma:2011}.

The first experiments on squeezing in atom interferometry showed that number squeezed states could extend coherence time of an interferometer \cite{Jo:2007aa,Li:2007aa}. Although these two experiments were different in details, the underlying principle is the same. A coherent state can be expressed as a sum over Fock states. In the presence of interactions, the phase of each term in the Fock state expansion develops at a different rate and an initially phase certain state evolves rapidly to a phase uncertain state, limiting the coherence time in an interferometer. In both experiments, repulsive interactions between the atoms produced states with sub-Poissonian number variance on beam splitting. The reduced number variance manifested as an increased coherence time in the experiment. In related work, under conditions well below the BEC transition temperature where interactions dominate, Maussang \etal\ directly observed sub-Poissonian number fluctuations 4.9\,dB below the shot noise limit after beam splitting in a magnetic trap on a chip. Close to the transition temperature, the fluctuations exceeded 3.8\,dB above the shot noise limit \cite{Maussang:2010fk}.

Recently, two groups have used the one-axis twisting scheme to produce spin squeezed states suitable for two-state interferometry beyond the classical precision limit. Riedel \etal\ demonstrated 3.7\,dB of squeezing in a superposition of the \ket{F=1,\,m_F=-1} and \ket{F=2,\,m_F=+1} states of \Rb{87} coupled via a two-photon microwave-rf transition on an atom chip \cite{Riedel:2010aa}. The inter-state interactions were controlled by means of state-dependent trapping potentials, which were used to enhance the Kerr nonlinearity. Spin noise tomography of the state allowed full reconstruction of the Wigner function of the spin-squeezed state, with a coherent spin squeezing factor of $\xi_S = 0.56$.

In a breakthrough experiment in 2010, the group of M. Oberthaler demonstrated the first atom interferometer with sensitivity greater than the classical limit \cite{Gross:2010aa}. 8.2\,dB of spin noise suppression was detected via spin noise tomography following a nonlinear beamsplitter employing the one-axis twisting mechanism. In this experiment, an interspecies Feshbach resonance was used to amplify the Kerr nonlinearity during the beamsplitting process. In a subsequent Ramsey interference experiment, fluctuations in the final atomic spin projection $J_z$ were measured to be below projection noise, affording a phase sensitivity better than the classical limit of $\Delta\phi = 1/\sqrt{N}$.

All the squeezing methods discussed here could find application in future precision measurements with BECs and atom lasers. However, most inertial atom interferometers do not operate at the projection noise limit, but rather at limits imposed by technical considerations. Additionally, the demonstrations of spin squeezing thus far have employed small numbers of atoms ($\sim10^3$) for which the shot noise limit is relatively high. Nonetheless, the creation of squeezing and entanglement in atomic beams has great potential in fundamental studies as well as in precision measurement, and will undoubtedly be pursued further in coming years.

\section{Pumped atom lasers and continuous operation} \label{pumping}

All atom lasers demonstrated to date have operated in a pulsed or quasi-continuous fashion, a restriction enforced by depletion of the source Bose-Einstein condensate. The demonstration of truly continuous, phase-coherent operation would open atom lasers to a host of new applications, and could extend the utility of atom-based inertial sensors to much higher bandwidth measurements. In this section, we review progress towards this outstanding goal.

\subsection{Comparison of cw optical and atom lasers}

In developing a continuous atom laser, it is useful to consider the design of the cw optical laser and the similarities and differences between bosonic atoms and photons. In vacuum, there are three important differences:
\begin{itemize}
\item Atoms interact with atoms and do not require a a separate gain medium to take advantage of Bose-stimulated scattering to produce a macroscopic population in the lasing mode. Photons do not interact with photons in vacuum, and thus require a gain medium with population inversion to mediate photon-photon scattering.
\item Atoms are conserved and any pumping process will require a reservoir of atoms in a storage trap. The reservoir will need to be replenished as atoms are pumped to the lasing mode. Photons are not conserved and a reservoir of photons is not necessary. 
\item Atoms are massive particles and display a quadratic dispersion relation, whereas photons are massless and display a linear dispersion relation as required by special relativity. There is no rest frame for a photon.
\end{itemize}
Despite these differences, optical lasers and atom lasers both rely on Bose-enhanced scattering to macroscopically populate a single trapped mode. Bose-enhanced scattering is a consequence of the requirement of symmeterisation of the many-body wavefunction: the transition rate into a mode already occupied by $N$ bosons is proportional to $N+1$ because there are $N+1$ contributions to the scattering process. In optics, this process is called stimulated emission. That the same physics is also present in atomic systems is on the one hand obvious, and on the other, quite fascinating.

To realize a continuous atom laser, the source Bose-Einstein condensate must be replenished or `pumped' in a sustained and phase-coherent fashion. There are three critical requirements that are necessary to realise a pumped atom laser:
\begin{enumerate}
\item Pumping of the lasing mode (matter-wave amplification) must proceed via Bose stimulated scattering.
\item The pumping mechanism must be irreversible in order to achieve a net transfer of atoms into the lasing mode. 
\item Pumping must be compatible with both continuous replenishment of the reservoir and continuous output-coupling from the lasing condensate.
\end{enumerate} 
Pumping (matter-wave amplification) that proceeds via a Bose-stimulated step ensures that the atoms pumped to the lasing mode enter with the appropriate phase, hence criterion 1 in the list above. Criterion 2 requires that one of the scattering partners in the pumping process leaves the system and does not return at least on the timescale of the experiment. In experiments of this kind, the only options are for an atom or a photon to leave. The third criterion is a consequence of non-conservation of atom number (in contrast to photons). An irreversible matter-wave pumping mechanism will require a geometry such that three processes can take place simultaneously: (i) replenishment of the reservoir, (ii) Bose-stimulated scattering (pumping) of the lasing mode,\footnote{According to early analysis, noisy pumping mechanisms may excite fluctuations in the lasing condensate, consequently broadening the linewidth \etal\ \cite{Haine:2002,Haine:2003}. Possible solutions to this problem include a spatially selective pumping scheme \cite{Johnsson:2005} or feedback control \cite{Haine:2004,Szigeti:2009,Szigeti:2010}.} and (iii) output-coupling from the lasing mode to produce a beam. Although no device to date has satisfied all of these criteria, there have been some important steps made along the path to this device.

\subsection{Matter-wave amplification}

\begin{figure}[t!]
\centerline{\scalebox{.8}{\includegraphics{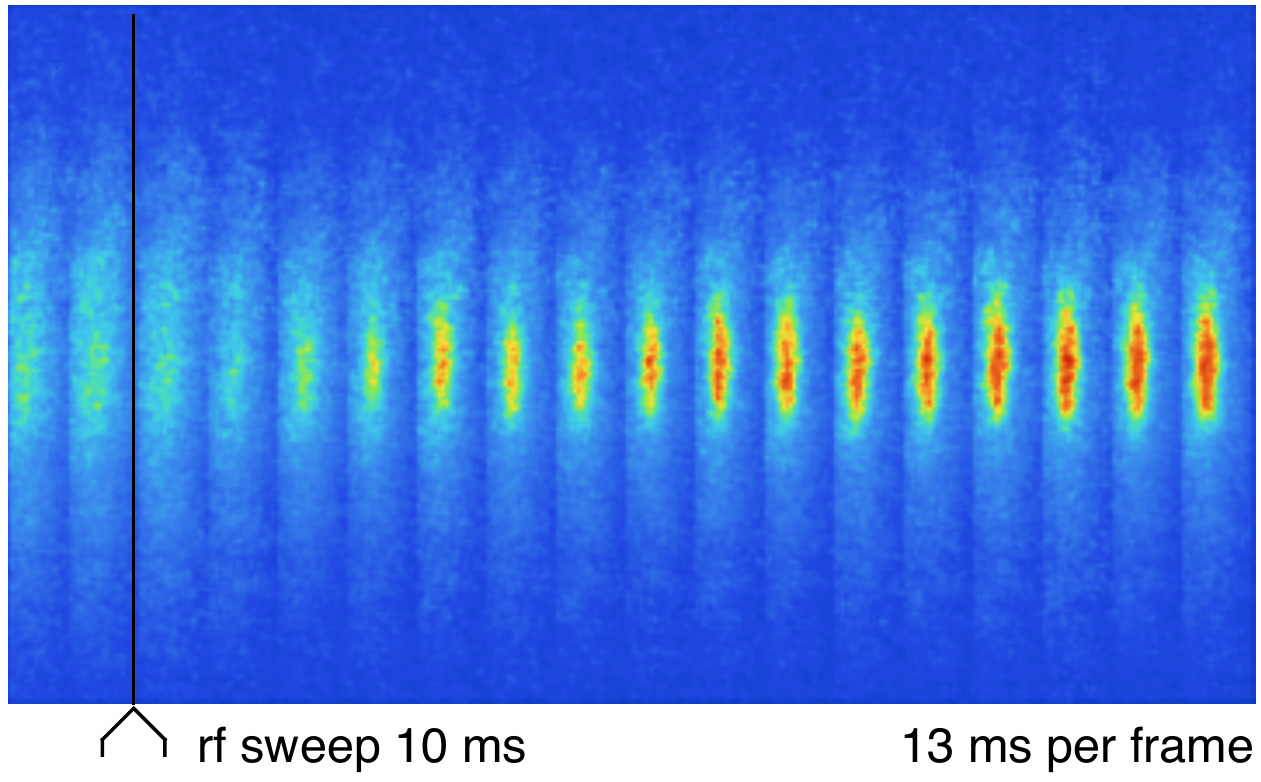}}}
\caption{The formation of a Bose-Einstein condensate via evaporation is an example of unseeded matter-wave amplification. Shown here is a sequence of 18 minimally-destructive phase-contrast images capturing the formation of a condensate in situ during a single experimental run. The first two frames show a thermal cloud at a temperature above the transition temperature. The following 16 frames were taken after the cloud was quenched to below the BEC transition and show the growth of a condensate at the centre of the cloud at 13\,ms intervals. Note the decrease in the number of thermal atoms and their smaller width after the rf sweep. The column density of atoms is shown in false colour: yellow to red marks the high density of the condensate. The length of each image is 630\,$\mu$m. Reproduced from \cite{Miesner:1998zr} with permission of the authors.}
\label{ketphasecont}
\end{figure}

The first realisations of atomic BEC in fact constituted the first unseeded matter-wave amplification experiments \cite{Anderson:1995vn,Davis:1995aa,Bradley:1995ys}. Evaporative cooling relies on binary collisions that lead to the loss of one of the scattering pair from the system and a change in population of the trapped modes. Under appropriate conditions, the scattering rate into the ground state increases with the number of atoms already in the ground state, and the population in the ground state grows exponentially in time as a consequence of Bose-stimulated scattering. This was directly observed by the MIT group using far detuned phase contrast imaging in 1999 \cite{Miesner:1998zr}. Images showing the formation of a condensate in a single experimental run are shown in Figure \ref{ketphasecont}.

%%%%%
%%%%%

In a conventional, un-pumped atom laser, the amplification process (evaporation) runs first to produce the condensate, then a beam is out-coupled. The two are not simultaneous, do not satisfy criterion 3 above, and the device is pulsed. Evaporation appears difficult to implement with continuous operation because the out-coupling process leads to heating. Out-coupling truncates the \emph{low} energy tail of the thermal distribution leading to heating and destruction of the condensate. It would seem resonable that with a suitably low out-coupling rate and strong truncation, evaporation and out-coupling could proceed together, forming the basis of a continuous atom laser. This parameter space has not yet been carefully investigated experimentally and may be a feasible path at very high pump, and low output-coupling, rates.

A variation on the standard evaporation geometry was investigated by Chikkatur \etal\ \cite{Chikkatur:2002dq}. In this experiment, condensates were produced by evaporation in a magnetic trap, transported using moving optical tweezers to an optical dipole trap and merged with a larger condensate. The merging of two condensates is not a Bose-stimulated process. As the authors discuss, merging two condensates with random phases will produce excitations that must be dissipated. It was argued that, provided the incoming condensate is substantially smaller than the main condensate, the phase of the larger BEC will dominate. Phase stability was not measured in this experiment, and the group did not out-couple an atom laser beam. In terms of development of a pumped atom laser, the most important contribution here was perhaps geometric. Pumping by evaporation was spatially separated from the lasing condensate, and it would appear to be possible to continuously out-couple an atom laser beam while simultaneously delivering fresh condensates produced by evaporation in a different location.

\subsection{Superradiance}

The alternative to evaporation is to exploit spontaneous emission to achieve irreversibility. This was extensively discussed early in the development of atom lasers in the theoretical literature \cite{Hope:1996ve,Williams:1998ly,Law:1998fk,Moore:1999vn,Bhongale:2000zr}. The basic idea is that an excited atom emits a photon and makes a transition to a highly occupied many-body state. In the process, the spontaneously emitted photon leaves the system to ensure irreversibility. The photon is spontaneous in the sense that it enters the photon vacuum field, but under conditions whereby the process is stimulated by a macrosocopically occupied matter-wave field (the lasing condensate). The challenge is to avoid heating of the lasing condensate due to reabsorption of the spontaneously emitted photons. 

The MIT group performed the first experiments demonstrating matter-wave amplification by spontaneous emission of a photon. The first mechanism investigated was Rayleigh superradiance, which bears a strong similarity to far detuned Bragg scattering \cite{Inouye:1999aa}. In a Bragg process, two phase-locked optical beams propagating in different directions are incident on an atom. The beams are far detuned from resonance with an excited state to suppress spontaneous emission, but have a frequency difference corresponding to the recoil energy of the atom. In a two-photon Bragg process, the atom absorbs a photon from the beam with higher frequency and is stimulated to emit a photon into the beam with lower frequency. In doing so it recoils to conserve momentum. The emission process is stimulated by the high occupation of the second optical field, and the system exhibits Rabi flopping with the atom cycling between the \ket{0\hbar k} and \ket{2\hbar k} momentum states, where $k$ is the effective single-photon recoil momentum.

In Rayleigh superradiance, the lower frequency optical Bragg beam is absent. A single far detuned beam is incident on a condensate, and atoms absorb a photon from this beam and recoil as they spontaneously emit a photon into a random direction. Although the emission process is spontaneous in the sense that the emitted photon enters an unoccupied mode of the radiation field (in contrast to the Bragg process), it can be stimulated by the matter-wave field itself under the right conditions. In subsequent scattering events, recoil of an atom into a momentum state that is already occupied by $N$ atoms will be enhanced by a factor of $N+1$. That is, the mode coupling scales as the square root of $N+1$, reminiscent of the stimulated emission in Bragg scattering. The difference is that it is the high occupation of a matter-wave field rather than a photonic field that stimulates the Rayleigh superradiance process. Unlike Bragg scattering or scattering into a cavity mode, the photon leaves the system, making the process irreversible.

\begin{figure}[t!]
\centerline{\scalebox{.5}{\includegraphics{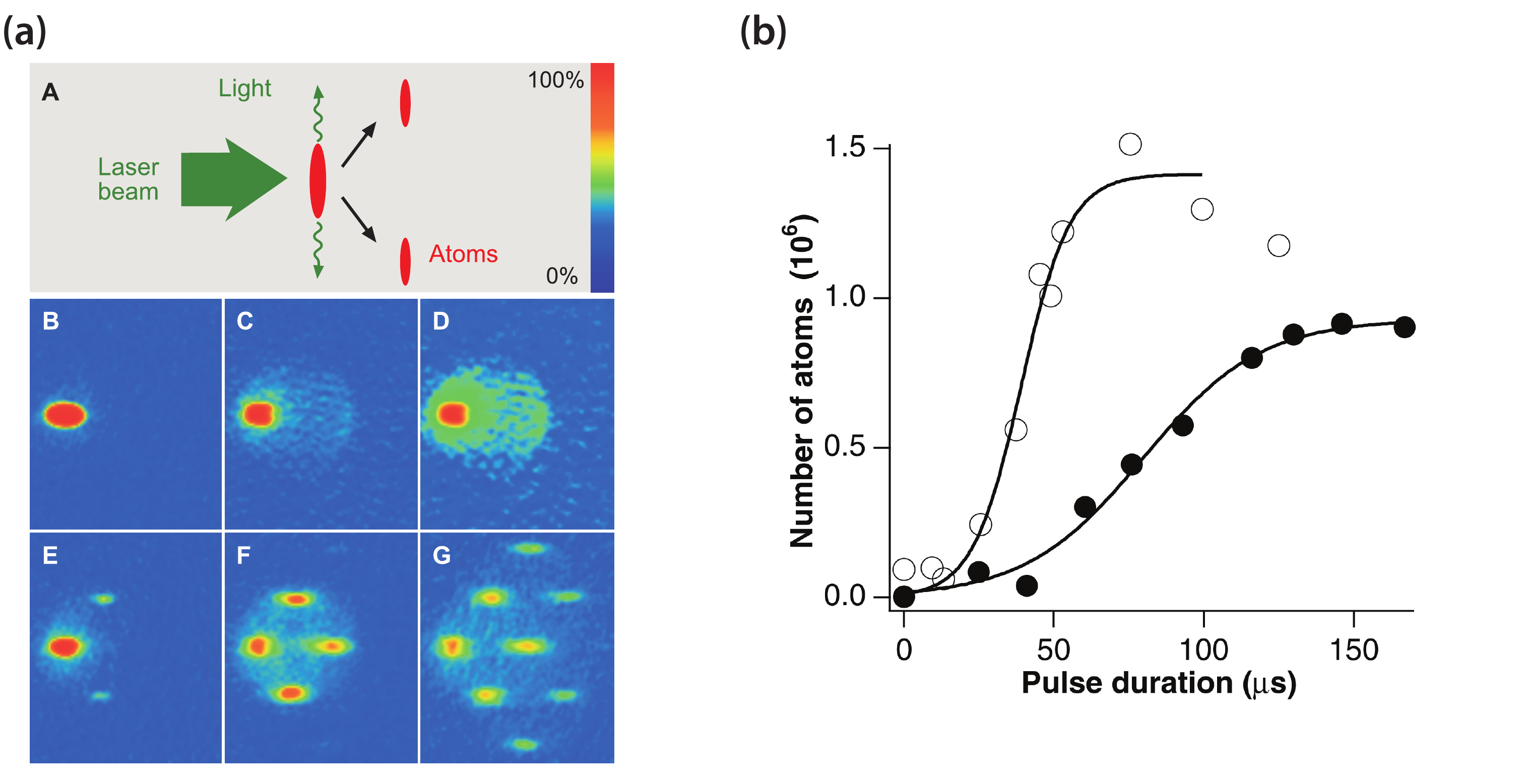}}}
\caption{(a) Observation of superradiant Rayleigh scattering. (A) An elongated condensate is illuminated with a single off-resonant laser beam. Collective scattering leads to photons scattered predominantly along the axial direction and atoms at 45$^\circ$. (B to G) Absorption images after 20\,ms time of flight show the atomic momentum distribution after their exposure to a laser pulse of variable duration. When the polarisation was parallel to the long axis, superradiance was suppressed, and normal Rayleigh scattering was observed (B to D). For perpendicular polarisation, directional superradiant scattering of atoms was observed (E to G) and evolved to repeated scattering for longer laser pulses (F and G). The pulse durations were 25 (B), 100 (C and D), 35 (E), 75 (F), and 100 (G) $\mu$s. The field of view of each image is 2.8\,mm by 3.3\,mm. The scattering angle appears larger than 45$^\circ$ because of the angle of observation. All images use the same colour scale except for (D), which enhances the small signal of Rayleigh scattered atoms in (C). (b) Observation of Òmatter-wave amplification.Ó Shown is the number of atoms in one of the superradiant peaks versus duration of the laser pulse. An intense atomic pulse was formed by amplification of spontaneous scattering. The initial number of atoms in the condensate at rest was $2\times10^7$, and the laser intensities were about 25 ($\bullet$) and 45 ($\circ$) mW/cm$^2$. The solid lines are guides to the eye. Reproduced from \cite{Inouye:1999aa} with permission of the authors.}
\label{superrad}
\end{figure}

In the Rayleigh superradiance experiment at MIT, mode competition selected very particular modes to be amplified, the so-called \emph{endfire modes}. The group observed an exponential increase in the number of atoms in the recoiling condensate as a function of the duration of the pump pulse (Figure \ref{superrad}b). At short pulse durations, two-photon processes dominate and the $2\hbar k$ boosted state alone is amplified. For longer pulse durations, superradiance populates a fan of momentum states (Figure \ref{superrad}a). Although this scenario is not ideal for pumping an atom laser, and the geometry of the experiment was not conducive to replenishment of the source (stationary) condensate, this work was the first demonstration of matter-wave amplification exploiting the emission of a photon to ensure an irreversible step. 

%%%%%
%%%%%

The early work on superradiance by the MIT group was unseeded and the phase of the amplified component of the matter-wave would have been random between different realisations of the experiment. Following this work, the group investigated seeded matter-wave amplification \cite{Inouye:1999vn}, using a Bragg pulse to seed the target state that was then populated by Rayleigh superradiance. The phase of the amplified state was then determined by the phase difference between the Bragg lasers and could be scanned in a Ramsey interferometer to measure phase stability of the amplification process. The MIT group performed just such an experiment to establish the phase coherence of the amplification process. An essentially identical experiment was performed contemporaneously by Kozuma \etal\ obtaining similar results \cite{Kozuma:1999aa}.

In later work, the MIT group investigated Raman superradiance \cite{Schneble:2004aa}, as did Yoshikawa \etal\ \cite{Yoshikawa:2004aa}. The basic physics is similar to Rayleigh superradiance. The important difference is that the initial and final states are in different hyperfine manifolds, as dictated by the polarisation of the pump field. Therefore, in Raman superradiance only one momentum state is populated. This process is therefore compatible with the requirements of a pumped, continuous atom laser.

\subsection{A pumped atom laser}

\begin{figure}[t!]
\centerline{\scalebox{.33}{\includegraphics{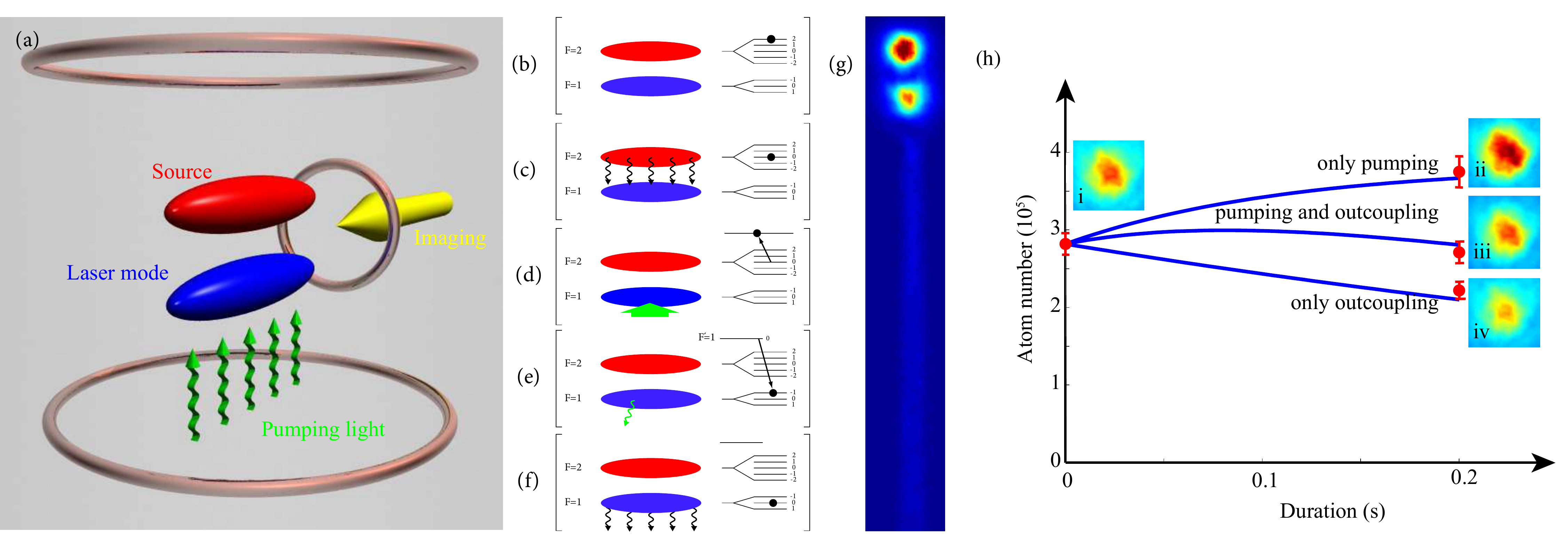}}}
\caption{Realization of a pumped atom laser, with simultaneous Bose-enhanced replenishment of and out-coupling from a lasing condensate. (a) Schematic diagram of the experiment and pumping steps (b-f). A radio frequency field spin flips atoms in the \ket{F=2,\,m_F=2} upper source condensate to the \ket{F=2,\,m_F=0} state (b) which then fall away under gravity (c). A light field propagating upwards couples atoms to the $F'=1$ excited state and they are stimulated to transition to the highly occupied \ket{F=1,m_F =- 1} mode of the lower lasing condensate. The momentum picked up in the 8\,$\mu$m fall between the two condensates is exactly cancelled by the absorption of the upward-propagating pump photon and emission of a downward-propagating photon (d,e). A second rf field out-couples atoms from the lower condensate to form the \ket{F=1,\,m_F =0} atom laser beam (g). (h) Atoms remaining in the laser mode after running the output-coupler and/or pumping. (i) No pumping or out-coupling (unpumped BEC), (ii) only pumping (pumped BEC), (iii) pumping and out-coupling (pumped atom laser), (iv) only output-coupling (un-pumped atom laser). The four data points are experimental data with error bars representing the standard deviation of the mean calculated from repeated independent measurements. The curves are calculated from a simple rate equation model that includes out-coupling as a loss and pumping as a gain in the number of atoms in the lasing mode. Reproduced from \cite{Robins:2008aa} with permission of the authors.}
\label{fig:robins}
\end{figure}

The ANU group demonstrated the first device capable of simultaneous Bose-stimulated pumping from a reservoir and out-coupling to produce an atom laser beam \cite{Robins:2008aa,Doring:2009aa}. In this experiment, two $^{87}$Rb condensates were prepared one directly below the other in the same magnetic trapping field (Figure \ref{fig:robins}). The upper condensate was in the \ket{F=2,\,m_F=2} hyperfine state and acted as the reservoir. The lower condensate situated 8\,$\mu$m below the reservoir was in the \ket{F=1,\,m_F=-1} state and was the lasing mode. Atoms were irreversibly pumped from the upper condensate to the lower condensate via a combination of rf coupling and Raman superradiance. The ANU group made a rate equation study of the atom laser clearly demonstrating simultaneous output-coupling from the lasing mode and Bose-enhanced pumping (Figure \ref{fig:robins}h). The phase coherence of the pumping process was not investigated in this experiment. Although the reservoir was not replenished, it was physically separated from the lasing atomic mode, and the geometry was compatible with replenishment of the reservoir. A truly continuous atom laser including replenishment of the reservoir remains a difficult and unrealised goal.

%%%%%
%%%%%

\section{Prospects for precision inertial measurements with atom lasers} \label{sect:precmeas}

This review has documented the development of atom lasers to date and has described their properties with specific reference to potential future applications in precision inertial measurements of gravity, acceleration and rotation. At the time of writing, all state-of-the-art cold atom inertial sensors are based on thermal sources. Whether this will remain so or whether the disadvantages of atom lasers can be overcome so that their advantages may be exploited is not yet clear. 

\subsection{The difference between clocks and inertial sensors}
Cold atoms, whether they are derived from thermal sources or atom lasers, provide universally available stable oscillators that can be applied to precision measurements of space and time. Atom-based technology revolutionised the measurement of time with the realization of the first atomic microwave clock in 1955. The principle of the measurement is straightforward. The phase of a countable microwave oscillator is written onto a stable atomic oscillator. The countable oscillator and the atomic oscillator are allowed to evolve independently, and their phases are compared after a period of free evolution. If the countable oscillator drifts in frequency during the free evolution time, the drift manifests as a phase shift that can be measured by an atom interferometer. The interferometer output is converted to an error signal and fed back to stabilise the countable oscillator. The cycles of the stabilised countable oscillator are the clock output. Such technology has allowed the creation of clocks that are stable to better than 1 part in $10^{14}$ in a 1 second measurement. 

The highest precision microwave clocks isolate the atomic oscillators from the environment by employing atoms in free fall that are derived from an atomic fountain. Although clocks and inertial sensors have many features in common, the application of atom lasers to clocks appears far less promising. All other factors being equal, the sensitivity of a microwave clock improves as the energy difference between the internal states coupled by the microwave radiation increases. This not true for inertial sensors. The sensitivity of an inertial sensor increases as the momentum imparted in the beam splitting process increases. The sensitivity (at least fundamentally) does not depend on the energy difference of the internal states that take part in the process. If a Bragg process is used, the beam splitters and mirrors couple different momentum states and only one internal state takes part in the process. The clock shift and interactions with stray fields that depend on the internal state of the atom are now common mode and Bragg-based inertial measurements can gain substantial immunity to effects of this kind. It is less important to operate at low density in a Bragg-based inertial sensor. The relative immunity of Bragg-based inertial sensors to the clock shift has been discussed by Debs \emph{et al.} \cite{Debs:2011aa,Debs:2012aa}. For clocks, however, it is essential to operate at low density. There appear to be few advantages and several disadvantages in using BECs and atom lasers for clocks. The only exception to this may be a high flux, low density, squeezed atom laser source.

Inertial quantities, broadly termed accelerations (but including gravity and rotations) can be measured using cold-atom technology similar to that employed in microwave clocks. Acceleration of the lab frame that carries the probing lasers is measured with respect to an inertial frame provided by freely falling cold-atom test masses. If the lab frame accelerates while the atoms are in free fall, the location of the atoms will change as measured by a ruler tied to the accelerating frame. The phase of the laser-ruler exhibits a strong spatial phase dependence evolving by $2\pi$ every half optical wavelength. Space is now encoded as phase, and the phase of the optical beams provide the ticks on the ruler. Again, it is an interferometric comparison of laser phase and atomic phase after a free evolution time that forms the basis of the measurement. A relative phase shift of lasers and freely falling atoms manifests as a population difference in the two output ports of the atom interferometer. The laser-ruler (Raman or Bragg beams) provide the beam splitters and mirrors. The relative population in the output ports encodes the phase shift and ultimately allows a measurement of the acceleration. The spatial mode and the momentum width of the atomic source are far more critical for inertial sensors than they are for clocks, and it is for this reason that atom lasers may find application to precision inertial measurement. The spatial mode of atom lasers and the transverse and longitudinal momentum width of the beam can be made orders of magnitude smaller than can be achieved with thermal sources of the same flux. 

\subsection{Large momentum transfer beam splitters and mirrors}

The fundamental sensitivity of a free space inertial sensor increases as the momentum imparted in the beam splitting process increases.  This has been discussed and investigated by a number of groups \cite{Denschlag:2002aa,Muller:2008aa,Muller:2009ab,Muller:2008ac,Clade:2009aa,Chiow:2011aa, Debs:2011aa}. This feature of an atom interferometer is arguably where the single most important practical advantage of an atom lasers may be found. The dramatically lower divergence, compared to thermal sources, afforded by a condensed source renders the beam splitting and mirror processes more efficient.   For example, recent work by Szigeti \etal\ investigated how to optimise Bragg pulses for finite momentum width atomic sources \cite{Szigeti:2012aa}. In particular, they found that the momentum width places a fundamental limit on the efficiency of Bragg mirrors, which in turn places a limit on usable flux (and hence signal-to-noise) at the output of the interferometer. 

Debs \etal\ have performed the only experimental study to date comparing the phase sensitivity of a thermal atom interferometer and a BEC or atom laser based interferometer in the same setup \cite{Debs:2011aa}. In their paper, Debs \etal\ presented results on a cold atom gravimeter using atoms derived from a BEC of \Rb{87}, and compared its performance to that achieved with a cold thermal sample in the same system. They observed an increase in fringe visibility when using a condensed source instead of a thermal one. Identical velocity selection pulses were applied before the interferometer, such that the projection of the atomic momentum distribution in the direction of the Bragg beams was identical for each case. However, the transverse momentum width of the two clouds differed by a factor of 3. In theory, for an ideal experiment this should not affect the sensitivity of the device. The condensed source showed an improved result compared to the thermal state, with the interference fringe visibility increasing from $(58\pm4)\%$ to $(85\pm11)\%$. Raising the temperature of the thermal state (broadening the transverse momentum width) led to decreasing fringe visibility. The authors speculate that expansion resulting from the broader transverse momentum spread of the thermal cloud made the measurement more sensitive to effects such as wavefront aberrations in the Bragg beams. The effect became more pronounced as the momentum transferred in the process was increased. Loss of fringe visibility with increasing momentum transfer and increasing interrogation time has been observed in every experiment of this kind performed to date. 

\begin{figure}[t!]
\centerline{\scalebox{.4}{\includegraphics{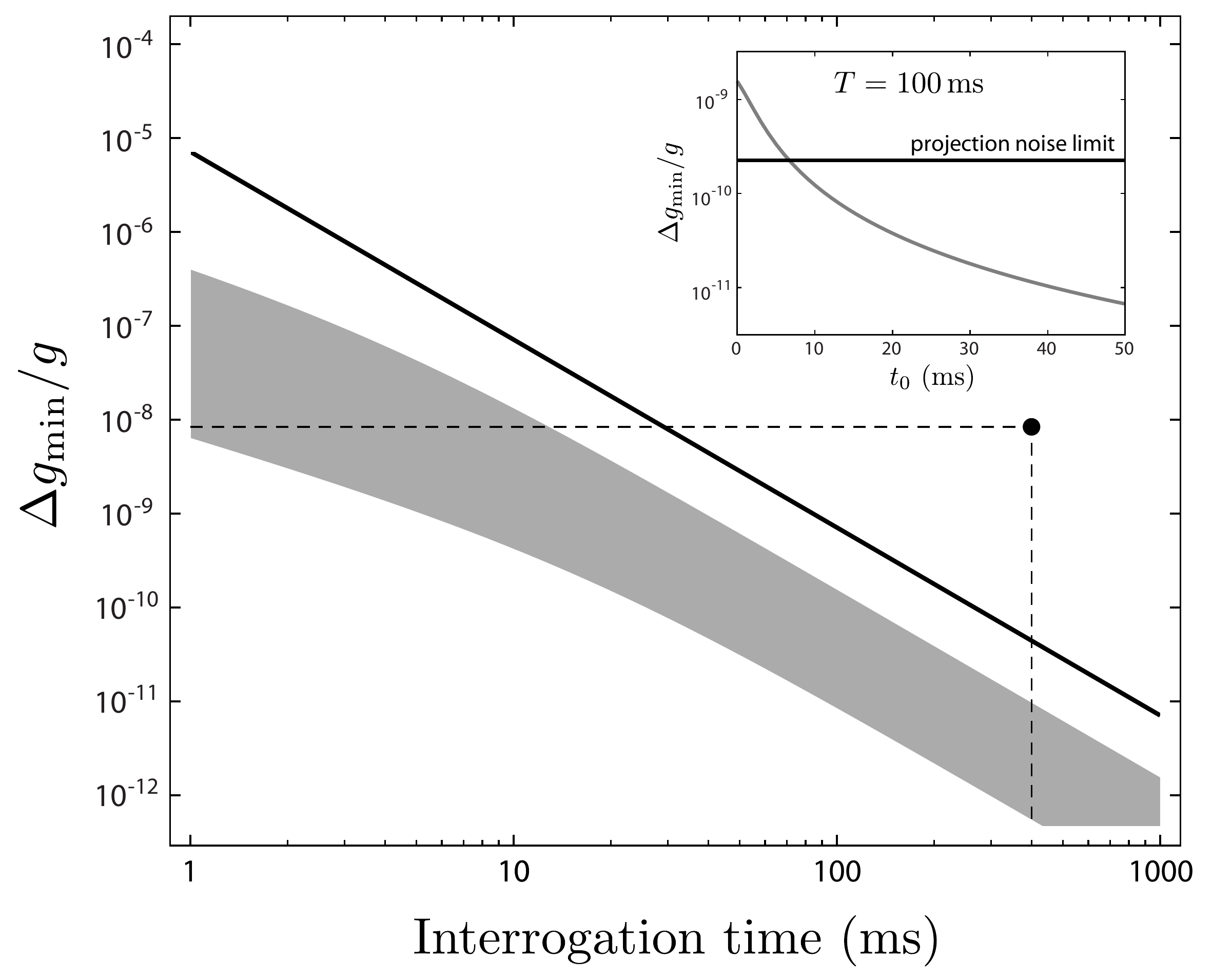}}}
\caption{The effect of interaction-induced phase diffusion on the sensitivity achievable in a BEC-based gravimeter. The shaded region shows the calculated diffusion-limited sensitivity for $N=10^6$ atoms and expansion times between 10\,ms (upper edge) and 50\,ms (lower edge). The solid line is the projection noise limit, and the dashed line is indicative of current state-of-the-art precision, obtained for a $2\hbar k$ momentum separation and interrogation time $T=400$\,ms \cite{Muller:2008b}. The inset shows the diffusion-limited sensitivity for $N=10^7$, $T=100$\,ms as a function of expansion time $t_0$. Increasing $t_0$ quickly reduces the effect of phase diffusion to below the projection noise limit.}
\label{fig:gravphasediffusion}
\end{figure}

In addition to the investigation of fringe visibility, Debs \etal\ theoretically examined the effect of atomic interactions in their free-space BEC gravimeter. The many-body quantum state of the condensate (initially in a coherent state) following the first beamsplitter pulse is distributed over states with different relative number. Each of these has a different interaction energy and thus acquires phase at a different rate, leading to a diffusion of the phase of the overall wavefunction \cite{Javanainen:1997,Castin:1997,Sinatra:2000} and thus a reduction in fringe contrast \cite{Widera:2008} (this is the same effect that can be exploited to generate squeezing -- see Section \ref{sect:squeezing}). While this effect can be significant in trapped atom interferometry, for inertial sensors for which the atomic test masses are in free-fall, the atom density is significantly reduced by ballistic expansion. An analytic estimate of the effect of interaction-induced phase diffusion was presented in Ref.\ \cite{Altin:2012aa} for a gravimeter based on a freely falling BEC. Figure \ref{fig:gravphasediffusion} shows the diffusion-limited sensitivity calculated from this estimate as a function of interrogation time $T$. The diffusion-limited sensitivity is well below the projection noise limit (solid line), and quickly drops below current state-of-the-art (dashed line) as $T$ is increased. Increasing the time allowed for ballistic expansion of the condensate before the interferometer also rapidly decreases the diffusion-limited sensitivity: the time $t_0$ allowed for ballistic expansion before the interferometer commences can always be increased to reduce the effect of interactions to below the projection noise limit, as shown in the inset. A more rigorous theoretical investigation of mean-field effects in a free-space interferometer has recently been published \cite{Jamison:2011}, and also concluded that atomic interactions do not preclude high precision measurements with atom lasers.

%%%%%
%%%%%

In the immediate future it is likely that BEC and atom lasers will be tested against traditional cold atom sources in high precision measurements of acceleration. Recent work has shown that a precision inertial sensor based on Bragg transitions of a thermal source can compete favourably with traditional Raman-based systems \cite{Altin:2012aaa}.  This work may lead to a direct comparison of BEC and thermal sources in the same apparatus at the limits of precision inertial measurement.    Experiments such as these will inform the design of future sensors, and likely settle the question of where, how and when atom lasers will make an impact on the field of precision measurement. A further important step forward will be the inclusion of high duty cycle BEC apparatus with ultra-high sensitivity inertial sensors.  Comparison tests of how LMT order and momentum width affect sensitivity will also likely be a focus for near term developments.

\section{Conclusion}

The conclusion that the classical properties of atom lasers such as brightness, low momentum width and coherence may be very useful in precision interferometry is consistent with experience in optics. In theory a laser and a light bulb of the same flux will provide the same signal-to-noise in an interferometric measurement of phase assuming they both operate at the shot noise limit. In practice, it is the classical properties of the laser that allow a myriad of technical difficulties to be overcome so that the device can operate at a high flux shot noise limit. It is for this reason that lasers are used in many precision optical measurements. Will the same be true for atom lasers? This remains an open question and one that should be pursued with vigour in the coming years. A continuous, pumped, squeezed atom laser suitable for precision measurement remains an important goal for the atom optics community.

\section*{Acknowledgements}

We gratefully thank Prof.\ Craig Savage, Dr.\ Joseph Hope, and Prof.\ Howard Wiseman for their inspiration, ideas, and discussions, particularly in the early days of our research program at ANU. We acknowledge the important contributions to our laboratory program at the ANU by Dr.\ Jessica Lye, Dr.\ Cameron Fletcher, Dr.\ Cristina Figl, Dr.\ Julien Dugu\'e, Dr.\ Matthew Jeppesen, Dr.\ Simon Haine, Dr.\ Graham Dennis, Dr.\ Mattias Johnsson, Mr.\ Stuart Szigeti, Mr.\ Gordon McDonald, Mr.\ Kyle Hardman and Mr.\ Shayne Bennetts. This work has been supported by the Australian Research Council through the Centres of Excellence and Discovery programs.

\section*{References}

\bibliographystyle{elsarticle-num}
\bibliography{alrev}

\end{document}